\documentclass[final,5p,times,twocolumn,authoryear]{elsarticle}

\usepackage{amssymb}

\usepackage{lineno}

\usepackage{graphics}
\usepackage[hidelinks,colorlinks=true,linkcolor=blue]{hyperref}
\usepackage{amsmath}
\usepackage{bbm}
\usepackage{color}
\usepackage{tabularx}

\definecolor{darkred}{RGB}{200,50,50}
\definecolor{darkgreen}{RGB}{150,201,150}
\definecolor{darkblue}{RGB}{150,150,251}
\definecolor{darkmagenta}{RGB}{211,121,221}
\definecolor{myorange}{RGB}{255,197,40}
\definecolor{mycyan}{RGB}{77,217,247}
\definecolor{mypink}{RGB}{234,72,196}
\definecolor{myyellow}{RGB}{234,234,42}
\definecolor{mygreen}{RGB}{123,234,45}
\definecolor{myblack}{RGB}{23,23,23}
\definecolor{grey}{RGB}{201,201,201}

\newcommand{\mr}[1]{\mathrm{#1}}
\newcommand{\pl}{p_\mr{li}}

\newcommand{\NN}{\mathcal{N}}
\newcommand{\RR}{\mathbb{R}}
\renewcommand{\d}{\mathrm{d}}

\renewcommand{\a}{\alpha}
\renewcommand{\b}{\beta}

\newcommand{\GG}{\Gamma}

\newcommand{\sg}{\sigma}

\newcommand{\Ex}{\mathbbm E}

\newcommand{\ytr}{\bar y} 
\newcommand{\Xtr}{\bar X} 
\newcommand{\ypr}{y^\ast} 
\newcommand{\Xpr}{X^\ast} 
\newcommand{\y}{y} 
\newcommand{\Y}{Y} 
\newcommand{\hlag}{h_\mr{lag}}

\newcommand{\lk}{p_\mr{like}} 
\newcommand{\pr}{p_\mr{prio}} 
\newcommand{\po}{p_\mr{post}} 
\newcommand{\pv}{q_\mr{ppv}} 
\newcommand{\pd}{p_\mr{ppd}} 
\newcommand{\pde}{\hat p_\mr{ppd}} 
\newcommand{\crps}{S_\mr{\!CRPS}} 

\DeclareMathOperator*{\argmax}{arg\,max}

\journal{Applied Energy}

\begin{document}

\begin{frontmatter}



\title{Bayesian Hierarchical Probabilistic Forecasting of Intraday Electricity Prices}


\author[1]{Daniel Nickelsen\corref{cor1}}
\ead{daniel.nickelsen@uni-a.de}


\author[1]{Gernot M\"{u}ller}

\cortext[cor1]{Corresponding author}

\affiliation[1]{organization={Computational Statistics and Data Analysis, Institute of Mathematics, University of Augsburg},
	city={Augsburg},
	country={Germany}}

\begin{abstract}
We address the need for forecasting methodologies that handle large uncertainties in electricity prices for continuous intraday markets by incorporating parameter uncertainty and using a broad set of covariables. This study presents the first Bayesian forecasting of electricity prices traded on the German intraday market. Endogenous and exogenous covariables are handled via Orthogonal Matching Pursuit (OMP) and regularising priors. The target variable is the IDFull price index, with forecasts given as posterior predictive distributions. Validation uses the highly volatile 2022 electricity prices, which have seldom been studied. As a benchmark, we use all intraday transactions at the time of forecast to compute a live IDFull value. According to market efficiency, it should not be possible to improve on this last-price benchmark. However, we observe significant improvements in point measures and probability scores, including an average reduction of $5.9\,\%$ in absolute errors and an average increase of $1.7\,\%$ in accuracy when forecasting whether the IDFull exceeds the day-ahead price. Finally, we challenge the use of LASSO in electricity price forecasting, showing that OMP results in superior performance, specifically an average reduction of $22.7\,\%$ in absolute error and $20.2\,\%$ in the continuous ranked probability score.
\end{abstract}




\end{frontmatter}


\section{Introduction and Motivation}
In April 2024 the monthly share of renewable electricity generation in the European Union reached 52.8\% (\href{https://www.energy-charts.info/charts/renewable_share/chart.htm?l=en&c=EU&year=2024}{energy-charts.info}), a new high and for the first time surpassing the milestone of having a majority of electrical energy produced by renewables in the EU. This milestone has already been reached March 2019 in Germany's energy generation, now approaching a yearly share of 60\% renewables. But the penetration of Renewable Energy Sources (RES) necessary for achieving the vision of sustainable energy production comes with tough challenges, as, opposed to other sources of energy, wind and solar energy production cannot be planned in advance, and mechanisms need to be in place to deal with the intermittency of RES in order to keep our energy systems stable.

With the introduction of an intraday trading system in 2006 by the European Power Exchange (EPEX Spot SE, or just EPEX) and its subsequent expansions \citep{Viehmann_state_2017}, a market-based mechanism has been created in which market participants contribute to a stable operation of power grids \citet{Koch_short-term_2019,Narajewski2022}. In the continuous intraday (CID) market, energy is traded under the pay-as-bid principle, where individual transactions can be executed up to 5\,min before delivery. Such a real-time trading system allows for last minute adjustments and thus minimizes imbalances in the power grid. In fact, the CID market is claimed to resolve the ``German balance paradox'', in which, despite the fast growth of RES in the energy mix, balancing needs are continuously decreasing \citep{Remppis_Influence_2015,Koch_short-term_2019}. One reason is that remaining imbalances can be very costly for market participants \citep{Narajewski2022}, constituting an incentive to instead use the intraday market for balancing \citep{Koch_short-term_2019}.

Since its implementation, the intraday market has gained popularity every year. In 2022, 134.6\,TWh of the total of 611.21\,TWh traded on EPEX was traded on the intraday market, a new all-time high (\href{https://www.epexspot.com/en/news/power-markets-deliver-transparent-price-signals-under-increased-supply-pressure}{epexspot.com}). However, to optimise its function as a market-based solution to prevent imbalances caused by increasing penetration of fluctuating RES, \mbox{(semi-)automated} tools allowing swift response to changes in the market 24/7 are necessary \citep{Koch_short-term_2019}. It stands to reason that forecasting of electricity prices on the CID market will play an important role in these developments, in particular probabilistic forecasts enabling proper risk management.

For this reason, a research focus at the interface between statistical learning and market economy concerning Electricity Price Forecasting (EPF) of (mostly the German) CID market has been growing in the last few years%
. A recent review on the EPF literature can be found in \citet{Maciejowska_forecasting_2023}, a review focusing on probabilistic EPF is \citet{Nowotarski_recent_2018}.

The majority of works on CID EPF concerns volume-weighted averaged prices (VWAPs) of executed transactions on the CID market. A few works instead focus on other aspects: Instead of only considering executed transactions, \citet{Shinde2021,Scholz2021} investigated the whole order book of the CID market. Arrival times of trades in the CID market was studied in \citet{Narajewski2019}. Since the role played by the balancing system is similar to that of the CID market, \citet{Narajewski2022,Lima_bayesian_2023} have focussed on forecasting imbalance prices.

\begin{figure*}[t!]
	\begin{center}
		\includegraphics[width=0.85\textwidth]{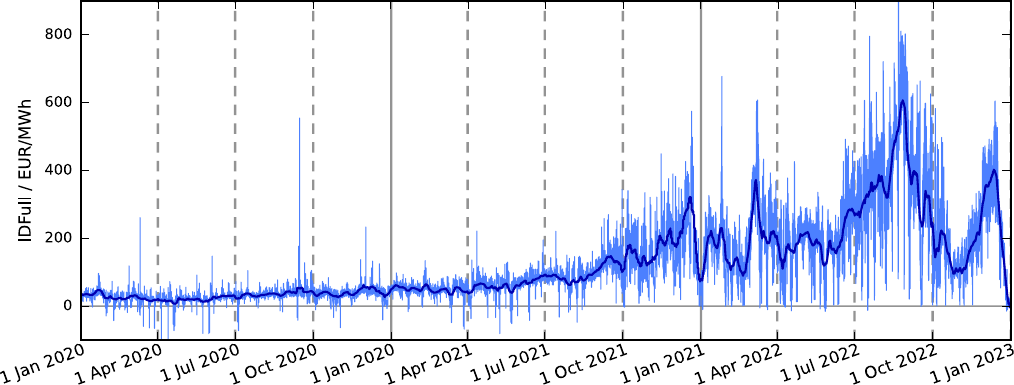}
	\end{center}
	\caption{The unprecedented increase of volatility in the (German) intraday electricity market is illustrated by the volume-weighted average price (IDFull) for all hourly products for 2021 and 2022 shown as a thin line. The thick line is a weekly rolling average. The data was provided by \citet{EPEXdata}.}
	\label{fig:VWAPs}
\end{figure*}

Point forecasts of electricity prices are typically based on regularised regression models \citep{Kremer_intraday_2020,Kath_value_2018}, in particular autoregressive models \citep{Hu_effects_2021,Lucic_performance_2023}, as well as artificial neural networks \citep{Oksuz2019,Lehna_reinforcement_2022}, or, more in the context of electricity demand forecast, functional data approaches \cite{vilar_forecasting_2012,shah_functional_2022,varelas_forecasting_2024}. 

However, in view of the strong volatility in the CID market, in particular in Germany with their high solar and wind energy components in the energy mix (cf. Figure \ref{fig:VWAPs}), most of the recent works on CID EPF agree upon the importance of probabilistic forecasting \citep{Maciejowska_forecasting_2023}. Employed probabilistic forecasting methods include generalised additive models for location, scale and shape (GAMLSS) \citep{Abramova2020,Narajewski2020,Narajewski2022,Hirsch2022,Hirsch_multivariate_2023}, distributional neural networks (DDNNs) \citep{Marcjasz2022,Narajewski2022,Barunik_learning_2023}, quantile regression averaging (QRA) \citep{Marcjasz2022,Maciejowska_probabilistic_2023,Andrade2017,cabrera_forecasting_2017,Uniejewski_regularized_2021,Serafin_averaging_2019}, and other specialised techniques \citep{Cramer_multivariate_2023,Grothe_point_2023}. However, all of these approaches use point estimates of distribution parameters (or quantiles) to build a probabilistic forecast model, and as such do not account fully for the uncertainty of parameter estimation. With our work, we fill this gap by promoting a Bayesian forecast model tested on the German CID market for the recent period of 2021 - 2022 featuring particularly volatile electricity prices.

Only very few works exist that employ fully Bayesian methods for EPF. An early example is \citet{Panagiotelis_bayesian_2008} for the Australian intraday market with a test set of 30 days in 2006. A more recent application, again to the Australian intraday market, use Bayesian recurrent networks \citep{Klein_deep_2023}. In Europe, the recent work \citet{Brusaferri_bayesian_2019} applies Bayesian deep learning to the Italian and Belgian day-ahead market. Here, the intractable posterior is approximated by a Gaussian model in a variational Bayes approach. Very recently, the British imbalance system has been subject of \citet{Lima_bayesian_2023}, in which through Bayesian updating and conjugate priors a semi-analytic Bayesian dynamic learning model with time varying parameters has been proposed. It is important to note that none of these Bayesian examples tackle the European intraday market.

Furthermore, none if the above mentioned examples tackle data later than 2019. In fact, of all literature on CID EPF in Europe, to the best of our knowledge, \citet{Hirsch_multivariate_2023} is the only work taking up this challenge and presenting a forecast study of German CID prices of the recent year 2022. In this work, the authors fit a mixture distribution involving Johnson's SU distribution and copulas to model all 24 hourly VWAPs of a day with their full dependency structure in a multivariate probabilistic model.

A crucial step of EPF is the choice of regressors. Typical regressors found in most works are price and volume information of the electricity markets, external variables like forecasted power generation of various energy sources and load forecasts, and dummy time and date variables to capture seasonal trends. But also carbon emission allowances \citep{Pape2016,Marcjasz2022,Maciejowska2020,Maciejowska_probabilistic_2023}, unavailability of power generation \citep{Pape2016,Hirsch2022,Lima_bayesian_2023}, cross-border flow of energy \citep{Pape2016,Andrade2017}, trade closing times \citep{Hirsch2022,Hirsch_multivariate_2023}, market elasticity \citep{Kremer_econometric_2021,Hirsch2022,Hirsch_multivariate_2023}, and grid frequency \citep{Scholz2021} have been utilised. The market elasticity is approximated by the slopes of supply and demand curves through two-sided auctions in the day-ahead market \citep{Kulakov_determining_2019}. Typically, also autoregressive components are considered as additional regressors \citep{Maciejowska2019,Maciejowska_enhancing_2021,Janke2019,Kath2019,Maciejowska2020,Uniejewski_regularized_2021,Uniejewski2019}. These ARX-type models typically use lags of one day or one hour. Often, a lag of two days and, to account for weekly seasonality, seven days is also considered, but these lags have been reported as less relevant \citep{Pape2016,Shinde2021}.

In view of this large set of regressors, possibly with strong collinearities, selection of regressors or features is crucial. The Least Absolute Shrinkage and Selection Operator (LASSO) \citep{tibshirani_regression_1996} is a popular choice and has been declared the gold standard of EPF \citep{Maciejowska_forecasting_2023}. In this work, referring to the known instabilities the LASSO has when facing strong collinearities \citep{Su2017}, we challenge this standard and instead promote Orthogonal Matching Pursuit (OMP) \citep{tropp_signal_2007,rubinstein_efficient_2008,pedregosa_scikit-learn_2011} as a feature selection technique for the case of large sets of regressors with strong correlations.

A recently debated topic is the efficiency of the European CID market. Evidence has been put forward, that the CID market is of weak-form efficiency \citep{Narajewski_econometric_2020}, picked up by \citet{Narajewski2020,Hu_effects_2021,Hirsch2022,Hirsch_multivariate_2023}, which, in essence, means that the last price information already contains all important information for the forecast of future prices. This claim, however, has been challenged in \citet{Marcjasz2020}. Here, we use the last price information as a benchmark model, and demonstrate that a small but statistically significant improvement is possible using our approach. But even though substantial improvements over last price information might be difficult to achieve, complementing the point information probabilistically still is of great value for optimal trading \citep{Kath_optimal_2020,Uniejewski_regularized_2021}.

In this work, we focus on the electricity traded on the EPEX CID markets in Germany in the years 2021 and 2022. We take the perspective of energy producers who, at a certain point of time $\tau$, are interested in CID electricity price levels in all later hours $h$ of the considered delivery day $d$. In view of the presumed weak-form efficiency of the CID market, and the unprecedented volatility in Germany for the years considered in this study, we propose a Bayesian probabilistic model which fully incorporates uncertainties and as such does not require fitting or training. The basic model structure is of ARX-type, selection of regressors is performed by OMP based on a large set of regressors. Additional regularisation is induced by our choice of prior. The forecasted price distribution is the Posterior Predictive Distribution (PPD) which we determine numerically by sampling the posterior with the No-U-Turn (NUTS) sampler \citep{Hoffman2014no}. Exploiting the probabilistic nature of our research, we compute probabilities for negative and positive price spreads, which is particularly relevant for practitioners \citep{Maciejowska2019,Maciejowska_enhancing_2021}.

Overall, our research can be broken down into the following contributions  to the field of EPF:
\begin{enumerate}
	\item To the best of our knowledge, we present the first complete Bayesian treatment of CID EPF, fully incorporating uncertainties of model parameters. The study period is the exceedingly volatile years 2021 and 2022, which, apart from the recent work by \citet{Hirsch_multivariate_2023}, has not been the subject of CID EPF before as far as we are aware.
	\item We address the problem of feature selection and present statistically significant evidence that OMP leads to a better forecasting performance than the declared gold standard LASSO.
	\item We add to the discussion of the proposed weak-form efficiency of CID markets and share a detailed description of CID indices and statistics calculations which have become more intricate since 2021, and, as far as we are aware, have not yet been published in this detail.
\end{enumerate}

This paper is structured as follows. In Section \ref{sec:markets}, we introduce the DA and CID markets in Europe with a focus on Germany, discuss merit order slopes, and provide full details on calculations of CID indices and statistics from EPEX transaction data. Section \ref{sec:datamodel} explains our data model, including all features used and the employed feature selection techniques. The data model feeds into our Bayesian forecast model which we define in Section \ref{sec:fcmodel}. In Section \ref{sec:results} our procedure to extract prediction intervals from the predictive distributions produced by the Bayesian forecast model is introduced, and the point measures and probability scores to evaluate our results are specified. In this section, we also present and discuss our results. We conclude our work in Section \ref{sec:conclusion}.

\section{Electricity markets and market data}\label{sec:markets}
Understanding the characteristics of the European electricity market is crucial for establishing successful forecasting models. In this section, we therefore summarise the market details important for our data model. More details of the (German) short-term electricity market can be found in \citet{Viehmann_state_2017} and the references given below.

\begin{figure*}[t!]
	\begin{center}
		\includegraphics[width=0.9\textwidth]{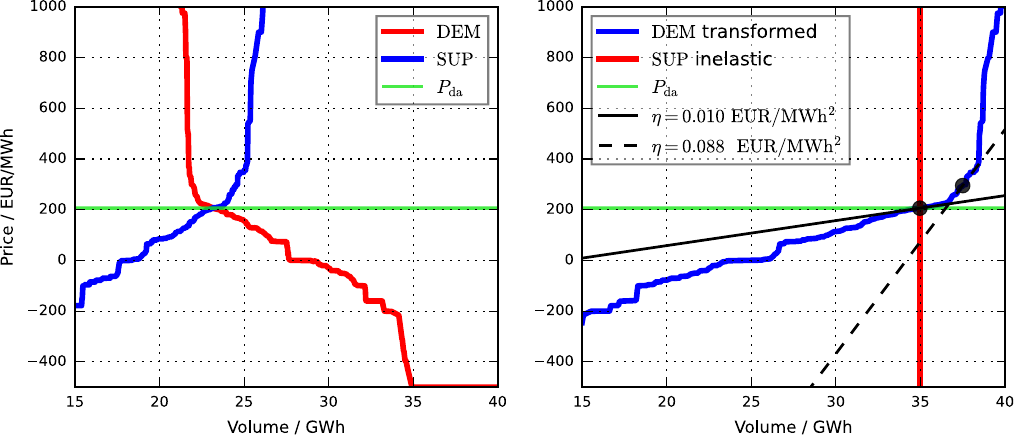}
	\end{center}	
	\caption{Illustration of the transformation proposed in \citet{Kulakov_determining_2019} to obtain the slope of the merit-order curve as a measure for elasticity $\eta$ in the DA market. The left side shows aggregated supply (SEP) and demand (DEM) curves as obtained from \citet{EPEXdata} for the delivery hour 7:00\,-\,8:00\,am on 4 November 2022, with the DA price $P_\mathrm{da}$ published in the European trading zone as a horizontal line. The right side depicts these curves after the transformation, in which the elasticity from the demand side is transferred to the supply side. Exemplary slopes are indicated by straight lines for two volumes marked by closed circles. A finite difference of 500\,MWh has been used to determine the slopes.}
	\label{fig:auctionslope}
\end{figure*}

The CID market resulted from the liberalisation of the European energy sector since 1996 (\href{https://www.europarl.europa.eu/ftu/pdf/en/FTU_2.1.9.pdf}{europarl.europa.eu}) and is one of many platforms to trade electricity. The most liquid market is the day-ahead (DA) market conducting a uniform price auction to settle clearance prices of electricity for the next day. Markets like the CID and DA markets in the European trading zone are operated by Nominated Electricity Market Operators (NEMOs), of which EPEX is one of 17 (\href{https://www.nemo-committee.eu/nemo_committee}{nemo-committee.eu}). However, due to the large share of volumes traded, EPEX is generally considered as the reference point for electricity prices in Germany and other countries \citep{Viehmann_state_2017}. In addition to NEMOs managing the trading platforms, various Transmission System Operators (TSOs) share the responsibility for the transmission of electrical power and for providing market participants access to the grid \href{https://www.entsoe.eu/about/inside-entsoe/members/}{(entsoe.eu)}.

In the following, we focus on the German market operated by EPEX. We will refer to the delivery day as day $d$, the day before delivery day as day $d-1$, and so on. Typically, electricity is traded in hourly, half-hourly and quarter-hourly delivery periods or \textit{products}. We will consider hourly products, that is 24 separate products per day, and a delivery hour denoted by $h$ will start at time $h$ and end 60\,min later, e.g. $h=14$ refers to the contractual period of 14:00\,-\,15:00\,pm. 

\subsection{Day-ahead market and elasticity}
The day-ahead market closes at 12:00\,noon on day $d-1$. Until then, a two-sided, blind auction for all products of day $d$ takes place, from which aggregated supply and demand curves are created. The intersection of these two auction curves defines the DA market clearance price and an associated DA volume, which is illustrated in the left graph of Figure \ref{fig:auctionslope}. Since the auction curves reflect the merit order on the spot market, it is also often referred to as \textit{merit-order} curves.

The slope of the supply curve can be used to measure the elasticity in the DA market, which may serve as a proxy for the elasticity in the CID market. However, due to the two sided auction, the elasticity of the demand curve should be taken into account as well. One way to accomplish that has been proposed in \citet{Kulakov_determining_2019}, in which a transformation transfers all elasticity of the demand side to the supply side, making the demand side perfectly inelastic. Now, the slope of the transformed supply curve at a certain volume measures the elasticity of the whole DA market for that volume. Figure \ref{fig:auctionslope} presents an illustration, \citet{Kulakov_determining_2019} gives full details, and the recent forecast studies \citet{Hirsch2022,Hirsch_multivariate_2023} employing this transformation for the CID market describe the method from a more practical perspective.

A subtle point in determining the DA market clearance price and volume is the different NEMOs and countries involved. Since 2014, the Single Day-Ahead Coupling (SDAC) creates a European trading zone for the DA market (\href{https://www.nemo-committee.eu/sdac}{nemo-committee.eu}) and unifies the determination of the market clearance price through an algorithm called Euphemia (\href{https://www.nemo-committee.eu/assets/files/euphemia-public-description.pdf}{nemo-committee.eu}). As a result, all participating countries and platforms, independent from the respective NEMO responsible, use the same SDAC price. The data published by EPEX since 2014 therefore includes only the unique SDAC price determined by Euphemia, but, on the other hand, is only authorised to publish the EPEX DA volume determined from EPEX auction curves in the respective country. From 14 October 2021, however, EPEX publishes the so-called All-Certified Exchanges' aggregated supply and demand curves determined from the SDAC auction \citep{EPEX}. The market clearance price and volume determined from the intersection of these curves will be the SDAC price and volume, while EPEX still publishes only their country-specific DA volume.

\subsection{Continuous intraday market and transaction data}
The CID market opens at 15:00\,pm on day $d-1$ for all products of day $d$. Similar to SDAC, the CID markets of participating countries in Europe are coupled through the Single Intraday Coupling (SIDC), initialised in June 2018 by NEMOs and TSOs (\href{https://www.nemo-committee.eu/sidc}{nemo-committee.eu}). On the SIDC platform, each product can be traded up to 60\,min before delivery starts, whereas trades within the same country and using the same NEMO platform on both sides only requires 5\,min lead time. The CID bidding on EPEX platforms takes place on the M7 trading system (\href{https://www.epexspot.com/en/tradingproducts}{epexspot.com}), where buy and sell orders can be placed, and as soon as two orders match, the transition is executed. Summaries and deeper analysis of the SIDC electricity market can be found in \citet{le_integrated_2019,Kath2019,Demir_exploratory_2020}.

These market details are important in order to work with trading data from these markets. The full order book is rarely analysed in the literature due to the vast amount of data and a considerable increase of noise from automated trading, examples include \citet{Shinde2021,Scholz2021}. In the present work, like in most of the CID literature, \citet{EPEXdata} provided us with the executed transactions of the German side of all EPEX operated trades. These transactions include all trades within EPEX Germany (sell and buy side), and all SIDC trades where the sell or buy side is using the EPEX platform in Germany. The most important data fields available per transaction for the purpose of this study is the volume traded in MWh (\texttt{Volume}) with the matched price in EUR/MWh (\texttt{Price}), trade identification number (\texttt{TradeId}), the trade execution time (\texttt{ExecutionTime}), the begin of the delivery period (\texttt{DeliveryStart}), the end of the delivery period (\texttt{DeliveryEnd}), a flag if the transaction takes the buy or sell perspective (\texttt{Side}), and a flag indicating whether the sell and buy sides are identical (\texttt{SelfTrade}). 

Commonly used aggregated information of CID trading are Volume-Weighted Average Prices (VWAPs), of which typical examples are the ID1, ID3 and IDfull price indices. The ID1 is the VWAP of all transactions that took place from 1\,h up to 30\,min before delivery start of a product, ID3 is the VWAP of all transactions within 3\,h up to 30\,min before delivery start, and the IDFull is the VWAP of all transactions of a product. The price indices are analysed in \citet{Narajewski_econometric_2020} in detail. 

\subsection{Reproduction of published EPEX CID price indices and statistics}
EPEX distinguishes prices statistics and price indices. Price indices, like the ID1, ID3 and IDFull, are always specified. If, for instance, a product has not been traded yet or with a volume below a certain threshold, a fall-back value is set, e.g. the spot price. A statistics of a product, on the other hand, only is defined if at least one trade has been made. The VWAP of a product, as the statistics ``Weight Avg.'' of this product, is, for this reason, only equal to the IDFull if the product has already been traded sufficiently\footnote{The difference between index and statistics becomes evident, if on (\href{https://www.epexspot.com/en/market-data}{epexspot.com}) the table view for the continuous market is selected, and the output for yesterday is compared with the output for today.}. Other statistics are the ``High'' and ``Low'', which reflect the largest and smallest price of all transactions of a product, respectively. The ``Last'' is the price of the most recent transaction for a product. 

The mentioned indices and statistics are officially published by EPEX (\href{https://www.epexspot.com/en/market-data}{epexspot.com}). With the complete transaction data available, we can reproduce the official values of these indices. However, care has to be taken to follow the exact same rules as EPEX does for accurate reproduction. As these rules are only vaguely specified on publicly available resources (\href{https://www.epexspot.com/sites/default/files/download_center_files/EPEX\%20SPOT\%20Indices\%202019-05_final.pdf}{epexspot.com}), and to the best of our knowledge have not yet been reported in the literature, we summarise the procedure that we found to be most accurate in reproducing the EPEX indices for the benefit of other researchers in CID EPF:

As a general rule, the following conditions must be satisfied for a transaction to contribute to EPEX price indices and statistics \citep{EPEX}.
\begin{enumerate}
	\item At least one side of the transaction is traded on a EPEX operated platform.
	\item The transactions need to have at least one side in the respective market area\footnote{In the EPEX transaction data, the market area is misleadingly given as \texttt{DeliveryArea}.} (here, Germany).
	\item The transaction has not been recalled or cancelled.
	\item The transaction is not indicated as being the result of a self-trade.
	\item The delivery start and end must match the product of interest (here, hourly products).
	\item Transactions listed with both sides in the data are counted only once.
\end{enumerate}
Criteria 1)-3) are met automatically by using the executed transaction data obtained from \citet{EPEXdata}. To ensure 4), we select transactions with \texttt{SelfTrade=N} (``No'') and \texttt{SelfTrade=U} (``Unknown''). The flag \texttt{U} can occur if one side of the transaction is on a trading platform operated by a different NEMO than EPEX \citep{EPEX}. Naturally, these transactions can only be SIDC trades\footnote{Note that not all SIDC trades need to be cross-border or cross-NEMO transactions, but, conversely, all cross-border or cross-NEMO transactions are the result of SIDC trades.}. Condition 5) is directly ensured by comparing (\texttt{DeliveryStart}) and (\texttt{DeliveryEnd}) with the desired product\footnote{Alternatively, one could use the \texttt{Product} information of transactions and select the products \texttt{Intraday\_Hour\_Power} and \texttt{XBID\_Hour\_Power}. However, occasionally, user-defined blocks of arbitrary delivery periods are traded which are listed as the closest product available, e.g. a 3\,h delivery period would still be a \texttt{Intraday\_Hour\_Power} product.}. Once the transactions are filtered to meet these conditions, duplicate entries have to be removed. The duplicates occur as EPEX includes all transactions executed on their platforms in the data, which includes the equivalent \texttt{BUY} and \texttt{SELL} sides if both sides are traded on the EPEX platform. To filter out these duplicates, we use the \texttt{TradeId} of each transaction, and only keep the transactions with the first occurring \texttt{TradeId}.

\subsection{Live intraday values}
The end-of-day (EOD) values of the CID indices and statistics are only available after gate-closure (i.e. 5\,min before delivery start after which no further trades are possible). As stipulated by the weak-form efficiency assumption of the CID market, the most recent transactions of a product carry the most relevant information for its EOD values. As these live transactions are also available to practitioners, it is imperative to make use of this information for EPF.

\begin{figure*}[t!]
	\begin{center}
		\includegraphics[width=0.9\textwidth]{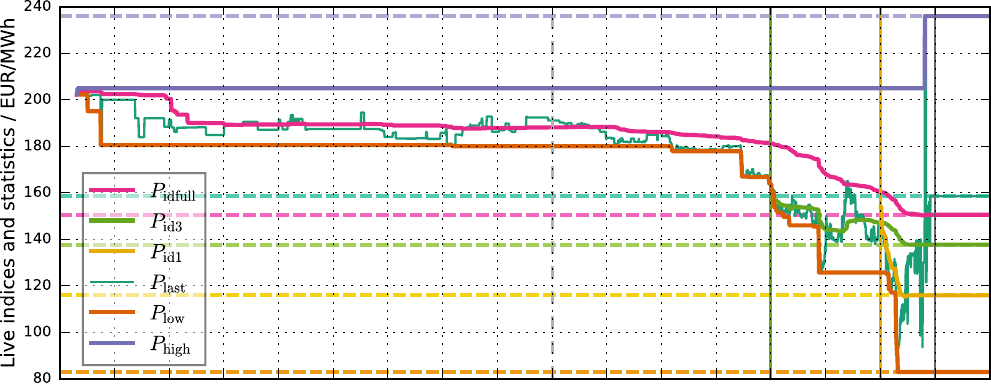}\\[1mm]
		\hspace{0mm}\includegraphics[width=0.9\textwidth]{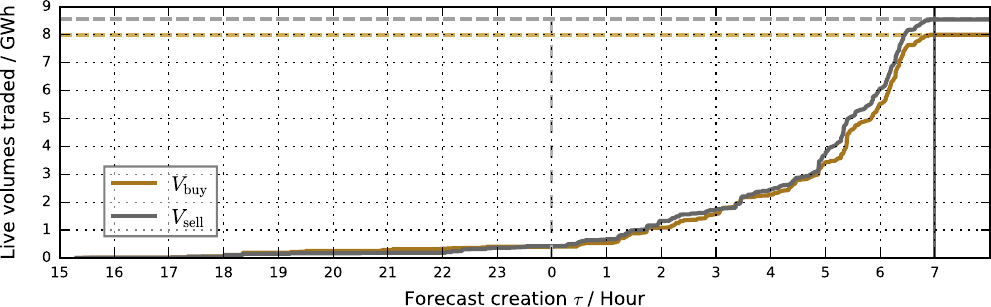}
	\end{center}	
	\caption{Live CID price and volume indices and statistics calculated from transaction data provided by \citet{EPEXdata} for the exemplary delivery hour 7:00\,-\,8:00\,am CET on 4 November 2022 as a function of forecast creation time $\tau$. The horizontal dashed lines indicate published values by EPEX. The dashed vertical line marks the beginning of delivery day $d$, i.e. hours left from this line are on day $d-1$ and on the right on day $d$. Solid vertical lines indicate 3\,h before delivery start and 1\,h before delivery start where applicable, and delivery start itself.}
	\label{fig:liveCID}
\end{figure*}

We therefore use the \texttt{ExecutionTime} information of transactions to take the perspective of a market participant trading power on the CID market at a thought point of time, and filter out any transactions that have taken place after this time. Thus, we emulate a forecast creation time $\tau$ and forecast the IDFull of all following hours. To indicate these preliminary values for CID price indices and statistics, we will use the prefix ``live'', e.g. \textit{live} IDFull.

For computationally efficiency, we pre-computed all live CID price indices and statistics for all products in 2021 and 2022 from EPEX transaction data, where $\tau$ varies on a dense time grid from 15:00\,pm on day $d-1$ to delivery end. We use linear interpolation to realise arbitrary values for $\tau$, alternative methods are discussed in \citet{Shinde2021}. Trying different numbers of grid points, we found 250 to be a good balance between resolution and computational cost. The inclusion of live market information into forecast models has only been picked up recently by a few works in the literature \citep{Marcjasz2020,Maciejowska2020,Hirsch2022,Maciejowska_probabilistic_2023,Hirsch_multivariate_2023}.

\begin{figure*}[t!]
	\begin{center}
		\includegraphics[width=0.9\textwidth]{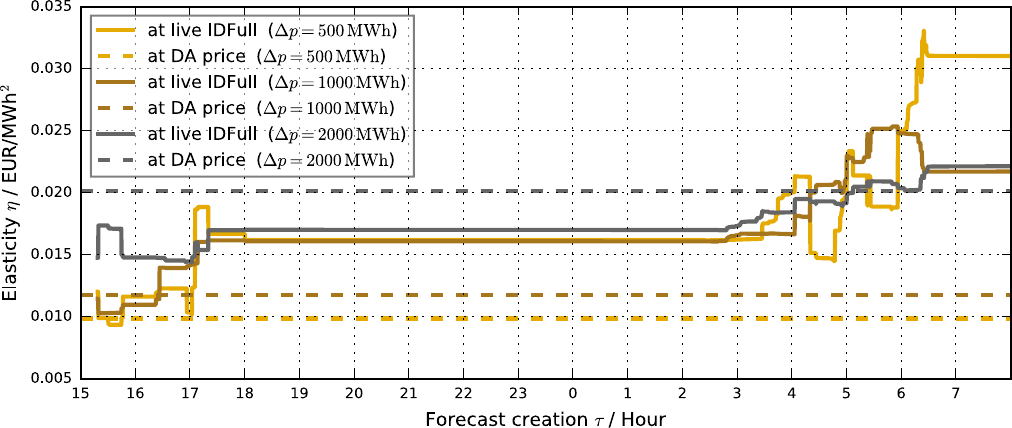}
	\end{center}
	\caption{Measures of elasticity of the DA market and the CID market using spot price and live IDFull for the exemplary delivery hour 7:00\,-\,8:00\,am CET on 4 November 2022. Here the three finite differences $\Delta p$ are used that are also used in the data model described in Section \ref{sec:datamodel}. The underlying live IDFull trajectory is shown in Figure \ref{fig:liveCID}, details on the elasticity estimation is illustrated in Figure \ref{fig:auctionslope}.}
	\label{fig:liveelasticity}
\end{figure*}

In Figure \ref{fig:liveCID}, we include an example to illustrate the generated data. We also added the published EPEX values to demonstrate their exact reproduction. Using the elasticity of the DA market as a proxy for the CID market, we can use the live IDFull to estimate the elasticity of the CID market along the same time line as illustrated in Figure \ref{fig:liveelasticity}.

\section{The data model}\label{sec:datamodel}
In this section, we describe the data model used to feed the Bayesian forecast model introduced in the next section. All variables listed are assumed to be for a fixed hour $h$, localised to CET or CEST depending whether or not daylight saving applies for the respective date and hour. If not stated differently, all variables are for delivery day $d$.

To manage clock changes, we adopt a no-clock-change rule in which the hour 3:00\,-\,4:00\,am is duplicated as the surrogate for the missing hour 2:00\,-\,3:00\,am in spring clock change, and of the twice occurring hour 2:00\,-\,3:00\,am in autumn clock change only the first occurrence is considered and the second is discarded. The rationale of this rule is that it gets as close as possible to a scenario without clock change, and hence captures the characteristic market dynamics of separate hours of the day with minimal distortion. 

\renewcommand{\arraystretch}{1.3}
\begin{table*}[t!]
	\begin{tabularx}{\textwidth}{lXl}
		Regressor & Definition & Source \\\hline\hline
		\multicolumn{2}{l}{Day-ahead market information \phantom{\LARGE I}} & \citet{EPEXdata} \\\hline
		$P_\mr{da}$ & Market-clearance price for SDAC countries (DA price, spot price). &  \\
		$P_\mr{da}^\mr{che}$ & Market-clearance price for Switzerland (non-SDAC). &  \\
		$V_\mr{da}$ & Market-clearance volume for EPEX Germany. &  \\
		$P_\mr{da}^{15,1}$, ..., $P_\mr{da}^{15,4}$ & The four quarter-hourly SDAC market clearance prices of delivery hour $h$. &  \\
		$s_{15}(P_\mr{da})$ & Average slope quarter-hourly SDAC market clearance prices, i.e. $s_{15}(P_\mr{da}) = \frac{1}{3}\Big(P_\mr{da}^{15,4} - P_\mr{da}^{15,1}\Big)$%
		& \\
		$\eta_\mr{da}(\Delta p)$ & Merit-order slopes at $P_\mr{da}$ calculated for three different finite differences $\Delta p=500,1000,2000$ as measure of elasticity. &  \\\hline %
		\multicolumn{2}{l}{Statistical day-ahead market information from other hours (\href{https://www.epexspot.com/sites/default/files/download_center_files/SFTP_specifications_2020-07.pdf}{EPEX moderated}) \phantom{\LARGE I}} & \href{https://transparency.entsoe.eu/}{ENTSO-E} \\\hline
		$P_\mr{da}(h_1,h_2,...)$ & Average DA price for selected delivery hours $h_1,h_2,...$ representing middle-night, early morning, morning, late morning, high noon, early afternoon, afternoon, evening, night, baseload, off-peak, rush hour, sunpeak, peakload, maximum, minimum, volume-weighted average. & \\\hline %
		\multicolumn{2}{l}{Continuous intraday market information \phantom{\LARGE I}} & \citet{EPEXdata} \\\hline
		$P_\mr{id1}$, $P_\mr{id3}$, $P_\mr{idfull}$ & Live CID price indices ID1, ID3 and IDFull. &  \\
		$P_\mr{high}$, $P_\mr{low}$, $P_\mr{last}$ & Live CID price statistics High, Low, Last. &  \\
		$P_\mr{deviat}$ & Live volume-weighted average deviation from mean price, i.e. %
		\begin{align*}
		P_\mr{deviat} = \frac{1}{\sum_{j\in I} v_j}\sum_{i\in I}v_i\,\Big[ p_i-\frac{1}{|I|}\sum_{j\in I}p_j\Big], 
		\end{align*}
		\mbox{where $\{(p_i,v_i)\,|\,i\in I\}$ are price and volume tuples of transactions executed before forecast creation.} &  \\
		$V_\mr{buy}$, $V_\mr{sell}$ & Live sums of volumes purchased and sold on the CID market.  &  \\
		$\mr{eod}_{d-1}(x)$ & \mbox{Final values of indices and statistics for day $d-1$,} \linebreak  \mbox{where $x=P_\mr{id1},\,P_\mr{id3},\,\y,\,P_\mr{high},\,P_\mr{low},\,P_\mr{last},\,P_\mr{deviat},\,V_\mr{buy},\,V_\mr{sell}$.} &  \\
		$\eta_\mr{cid}(\Delta p)$ & Merit-order slopes at $P_\mr{id1}$ calculated for three different finite differences $\Delta p=500,1000,2000$ as measure of elasticity. &  \\\hline %
	\end{tabularx}
	\caption{Base market regressors and their notation used in the model. If not stated differently, all variables pertain delivery day $d$ for a fixed hour $h$ in Germany. Forecasts and live values depend on availability at forecast creation time $\tau$. The background of these variables is explained in Section \ref{sec:markets}. External regressors are listed in Table \ref{tab:regress_base_ex}, additional features are calculated from these regressors and are listed in Table \ref{tab:regress_all}.}
	\label{tab:regress_base}
\end{table*}

\renewcommand{\arraystretch}{1.3}
\begin{table*}[t!]
	\begin{tabularx}{\textwidth}{lXl}
		Regressor & Definition & Source \\\hline\hline
		\multicolumn{2}{l}{Energy consumption and production forecasts for Germany \phantom{\LARGE I}} & \href{https://transparency.entsoe.eu/}{ENTSO-E} \\\hline 
		$E_\mr{cons}$ & \mbox{Energy consumption calculated as the average of day-ahead forecasts} of load for the four 15\,min periods of delivery hour $h$. &  \\ 
		$E_\mr{sol}, E_\mr{won}, E_\mr{woff}$ & \mbox{Solar, wind onshore and wind offshore energy production calculated as the} \mbox{average of the four 15\,min power forecasts of delivery hour $h$, where intraday} \mbox{forecasts replace day-ahead forecasts after 8:00\,am on day $d$.} & \\
		$E_\mr{tot}$ & {Total energy production calculated as the average of the four 15\,min day-ahead power forecasts of delivery hour $h$.} & \\
		$s_{15}(x)$ & \mbox{Average slope of 15\,min power forecasts, $x=E_\mr{cons},E_\mr{sol},E_\mr{won},E_\mr{woff}$, also see Table \ref{tab:regress_base}.} &  \\\hline %
		\multicolumn{2}{l}{Energy consumption and production forecasts for other countries \phantom{\LARGE I}} & \href{https://transparency.entsoe.eu/}{ENTSO-E} \\\hline
		$E_\mr{x}^\mr{sl}$ & \mbox{Day-ahead energy forecasts BSP Slovenia, $\mr{x} = \mr{cons},\;\mr{sol},\;\mr{won},\;\mr{woff},\;\mr{tot}$.} &  \\
		$E_\mr{x}^\mr{che}$ & \mbox{Day-ahead energy forecasts EPEX Switzerland, $\mr{x} = \mr{cons},\;\mr{sol},\;\mr{won},\;\mr{woff},\;\mr{tot}$.} &  \\
		$E_\mr{x}^\mr{itn}$ & \mbox{Day-ahead energy forecasts GME Italy, $\mr{x} = \mr{cons},\;\mr{sol},\;\mr{won},\;\mr{woff},\;\mr{tot}$.} &  \\
		$E_\mr{x}^\mr{hu}$ & \mbox{Day-ahead energy forecasts HUPX Hungary, $\mr{x} = \mr{cons},\;\mr{sol},\;\mr{won},\;\mr{woff},\;\mr{tot}$.} &  \\
		$E_\mr{x}^\mr{cz}$ & \mbox{Day-ahead energy forecasts OTE Czechia, $\mr{x} = \mr{cons},\;\mr{sol},\;\mr{won},\;\mr{woff},\;\mr{tot}$.} &  \\\hline %
		\multicolumn{2}{l}{Date and time dummy variables \phantom{\LARGE I}} & \href{https://docs.python.org/3/library/datetime.html}{\textsc{python} \texttt{datetime}} \\\hline
		$T_\mr{h}$ & \mbox{Hour of the day ($0-23$).} & \\
		$T_\mr{wd}$ & \mbox{Day of the week ($0-6$ for Mon-Sun).} & \\
		$T_\mr{m}$ & \mbox{Month of the year ($1-12$).} &  \\
		$T_\mr{y}$ & \mbox{Year ($2021$ or $2022$).} &  \\
		$T_\mr{wc}$ & \mbox{Weekday category, $0$ for Mon-Fri, $1$ for Sat, $2$ for Sun.} &  \\
		$T_\mr{deliv}=h-\tau$ & Time to delivery in floating point hours. &  \\
		$T_\mr{public}$ & Population-weighted number of states in work holiday. & \href{https://www.arbeitstage.org/}{arbeitstage.org} \\
		$T_\mr{school}$ & Population-weighted number of states in school holiday. & \href{https://www.ferienwiki.de/}{ferienwiki.de} \\\hline %
		\multicolumn{2}{l}{Market state variables} &  \\\hline
		$M_\mr{gas}$ & S\&P GSCI Natural Gas index. & \href{https://www.spglobal.com/spdji/en/indices/commodities/sp-gsci-natural-gas/}{S\&P Global} \\
		$M_\mr{gasoil}$ & S\&P GSCI Gasoil index. & \href{https://www.spglobal.com/spdji/en/indices/commodities/sp-gsci-gasoil/}{S\&P Global} \\\hline %
		\multicolumn{2}{l}{Daily and yearly seasonality indicators} & \href{https://sunrise-sunset.org/}{sunrise-sunset.org} \\\hline
		$S_\mr{tnoon}$ & Time to noon in floating point hours. &  \\
		$S_\mr{ttwi}^{1}$, $S_\mr{ttwi}^{2}$ & \mbox{Time to begin and end of twilight in floating point hours.} &  \\
		$S_\mr{dl}$ & Lengths of days as time from sunrise to sunset. &  \\
		$S_\mr{temp}$ & Yearly season indicator constructed from 15-year hourly temperature averages across weather stations in Germany. & \href{https://github.com/meteostat/meteostat-python}{meteostat} \\\hline %
	\end{tabularx}
	\caption{External base regressors and their notation, continuation of Table \ref{tab:regress_base}. The variables listed here are explained in Section \ref{ssec:datamodel_ex}.}
	\label{tab:regress_base_ex}
\end{table*}

\subsection{Market variables}
As market variables, we consider DA price $P_\mr{da}$ and volume $V_\mr{da}$, CID indices $P_\mr{id1}$, $P_\mr{id3}$ and $P_\mr{idfull}$, as well as the statistics $P_\mr{high}$, $P_\mr{low}$ and $P_\mr{last}$. We also consider the volume-weighted deviation to the mean price, $P_\mr{deviat}$, and the volume bought and sold on the CID market, $V_\mr{buy}$ and $V_\mr{sell}$, respectively. In addition to the live CID values of day $d$, we also add the end-of-day (EOD) values of these CID values for day $d-1$, which we denote by $\mr{eod}_{d-1}(x)$ for CID value $x$, e.g. $\mr{eod}_{d-1}(P_\mr{idfull})$ for the final value of the IDFull on the day before delivery day. 

Owing to the pan-European DA price, we only include the DA price of Switzerland. For Germany and SDAC countries, we also make use of DA price statistics provided by \citet{EPEXdata} that aggregate DA prices of various hours of the day, e.g. morning, night, rush hour, sun peak; details are published by \href{https://www.epexspot.com/sites/default/files/download_center_files/SFTP_specifications_2020-07.pdf}{EPEX}.

Apart from the hourly products, also the 15\,min and 30\,min products may contain market information relevant for EPF. Here, we consider the four 15\,min DA prices $P_\mr{da}^{15,1}$, $P_\mr{da}^{15,2}$, $P_\mr{da}^{15,3}$, $P_\mr{da}^{15,4}$ of hour $h$ as extra regressors, together with the average slope $s_{15}(P_\mr{da})$ of these prices, where $s_{15}(x)$ shall denote the 15\,min slope of the corresponding variable $x$.

Finally, we add DA and live CID market elasticities $\eta_\mr{da}(\Delta p)$ and $\eta_\mr{cid}(\Delta p)$ for finite differences $\Delta p=500$\,MWh, $\Delta p=1000$\,MWh and $\Delta p=2000$\,MWh, as proposed in \citet{Hirsch2022}.

We summarise these market base regressors in Table \ref{tab:regress_base}.

\subsection{External regressors}\label{ssec:datamodel_ex}
Apart from the above market variables, also external factors have an important impact on the price formation. These fundamental variables, according to the weak-form efficiency hypothesis of CID markets \citep{Narajewski_econometric_2020}, may already be incorporated into most recent prices. We still include a set of extra regressors for two reasons. Firstly, to shed some more light on the validity of the hypothesis, and secondly, to enable forecast uncertainty extraction from this data. 

Standard external variables are power generation and load forecasts publicly available from the transparency platform of the European Network of Transmission System Operators for Electricity (ENTSO-E) (\href{https://transparency.entsoe.eu/}{entsoe.eu}). These include day-ahead forecasts, i.e. created 18:00\,pm on day $d-1$, of power generation from a multitude of different energy sources, as well as load forecasts. In addition, intraday forecasts, i.e. created 8:00\,am on day $d$, are published for offshore and onshore wind power and solar power. Some forecasts pertain 15\,min periods, in these cases we aggregate all power forecasts to hourly energy productions, as well as to average slopes of 15\,min power forecasts when available. Depending on what is available at forecast creation, we pick day-ahead or intraday forecasts. 

Overall, we have hourly consumption $E_\mr{cons}$ derived from load forecasts, and hourly renewable energy productions comprising solar $E_\mr{sol}$, onshore wind $E_\mr{won}$, and offshore wind $E_\mr{woff}$, as well as the total hourly energy production $E_\mr{tot}$ from power forecasts. Using the notation $s_{15}(x)$ for the 15\,min slope of the corresponding variable $x$, we also consider $s_{15}(E_\mr{cons})$, $s_{15}(E_\mr{sol})$, $s_{15}(E_\mr{won})$ and $s_{15}(E_\mr{woff})$. To also take cross-border effects on price formation into account, we include power generation forecasts of Slovenia, Switzerland, Italy, Hungary and Czechia. 

Another typical set of regressors for EPF are date and time dummy variables. Here, we use hour of the day $T_\mr{h}$ ($0-23$), day of the week $T_\mr{wd}$ ($0-6$ for Mon-Sun), month of the year $T_\mr{m}$ ($1-12$), and the year $T_\mr{y}$. Additionally, we add a weekday category $T_\mr{wc}$ which takes the value $0$ for Mon-Fri, $1$ for Sat and $2$ for Sun, the time difference $T_\mr{deliv} = h-\tau$ between forecast creation and start of delivery (time to delivery), a measure $T_\mr{public}$ for the proportion of the German population in public holiday, and, similarly, a measure $T_\mr{school}$ for school holidays. 

An important aspect for EPF, especially for the CID market in recent years, are market states varying with time. Events strongly affecting the markets in 2021 and 2022 obviously arose in the course of the COVID-19 pandemic and the Russian invasion of Ukraine, but also smaller events like interest policies and legislations related to energy can play a major role. These changing market states impede training of forecast model with historical data as stationary dynamics is not given. In particular the years 2021 and 2022 show extreme fluctuations due to major changes in market states, cf. Figure \ref{fig:VWAPs}. Ideally, custom variables capturing all dimensions of market states may be incorporated as regressors, but in view of the difficulty of this task, we take the S\&P GSCI Natural Gas and S\&P GSCI Gasoil indices $M_\mr{gas}$ and $M_\mr{gasoil}$ as approximations for market states. These index values are of daily resolution and exclude weekends and holidays, we therefore always take the most recent available value with respect to delivery day and forecast creation.

Finally, seasonality strongly influences the price formation in electricity markets, where hourly, daily, weekly and monthly patterns can be observed. Apart from the dummy variables, and by considering each hour of the day separately, we also conclude a number of seasonality variables that capture different aspects of these patterns. Using daylight data from \href{https://sunrise-sunset.org/}{sunrise-sunset.org} we use time to noon $S_\mr{tnoon}$, time to begin of twilight $S_\mr{ttwi}^1$ and end of twilight $S_\mr{ttwi}^2$, and the time $S_\mr{dl}$ between sunrise and sunset (day length). Additionally, we employ average temperatures in Germany from the past 15 years to construct an indicator $-1\leq S_\mr{temp}\leq1$ for yearly seasons. With $S_\mr{dl}$ and $S_\mr{temp}$ we hence have two variables capturing yearly seasonality in terms of daylight and temperature, respectively. 

We summarise these external base regressors in Table \ref{tab:regress_base_ex}.

\subsection{Construction of feature space for regression}
Using the described base regressors, we construct a feature space spanned by these regressors and linear combinations of those. While meaningful combinations may be learnt automatically from the data, defining them manually allows their consideration in feature selection procedures before feeding the features into the forecast model.

Apart from the features valid for the delivery period of the product we are forecasting, also past values have an impact on the price formation, as is evident from the number of ARX-type models found in the literature \citep{Janke2019,Kath2019,Maciejowska2019,Uniejewski2019,Maciejowska2020,Maciejowska_enhancing_2021,Uniejewski_regularized_2021}. Here, we focus on the difference for delivery hour $h$ on day $d$ to the previous delivery hour, as well as the difference to the same hour $h$ on the previous day $d-1$. We denote by $\Delta_h(x)$ the difference of a feature value $x$ to the previous hour, and the difference to the previous day by $\Delta_d(x)$. For computing $\Delta_h(x)$, we use the same forecast creation time $\tau$ for both delivery hours, while for $\Delta_d(x)$ we also shift $\tau$ to day $d-1$ in order to ensure a common ground for the difference. We apply $\Delta_h(x)$ and $\Delta_d(x)$ to all features, tripling the feature space. 

\renewcommand{\arraystretch}{1.3}
\begin{table*}[t!]
	\begin{tabularx}{\textwidth}{lX}
		Regressor & Description \\\hline\hline
		\multicolumn{2}{l}{DA market and live CID market \phantom{\LARGE I}} \\\hline
		$V_\mr{cid} = (V_\mr{buy}+V_\mr{sell})/2$ & \mbox{Average of live volumes purchased and sold in CID market.} \\
		$\y - P_\mr{da}$ & Price spread between live IDFull and DA price. \\
		$V_\mr{cid} - V_\mr{da}$ & Volume spread between live CID volume and DA volume. \\
		$\eta_\mr{cid}(\Delta p) - \eta_\mr{da}(\Delta p)$ & Elasticity spread between CID and DA market for three different finite differences\linebreak $\Delta p=500,1000,2000$. \\\hline %
		\multicolumn{2}{l}{Energy production forecasts in reference to DA market and live CID market \phantom{\LARGE I}} \\\hline
		$E_\mr{res} = E_\mr{tot} - E_\mr{sol} - E_\mr{won} - E_\mr{woff} $ & Residual energy production \\
		$E_\mr{tot} - V_\mr{da}$ & DA excess energy production. \\
		$E_\mr{res} - V_\mr{da}$ & DA excess residual energy production. \\
		$E_\mr{tot} - V_\mr{cid}$ & CID excess energy production. \\
		$E_\mr{res} - V_\mr{cid}$ & CID excess residual energy production. \\\hline %
		\multicolumn{2}{l}{Energy consumption forecasts in reference to DA market and live CID \phantom{\LARGE I}} \\\hline
		$C_\mr{res} = E_\mr{cons} - E_\mr{sol} - E_\mr{won} - E_\mr{woff} $ & Residual energy consumption \\
		$E_\mr{cons} - V_\mr{da}$ & DA excess energy consumption. \\
		$C_\mr{res} - V_\mr{da}$ & DA excess residual energy consumption. \\
		$E_\mr{cons} - V_\mr{cid}$ & CID excess energy consumption. \\
		$C_\mr{res} - V_\mr{cid}$ & CID excess residual energy consumption. \\\hline %
		\multicolumn{2}{l}{Energy production forecast shifts \phantom{\LARGE I}} \\\hline
		$E_\mr{sol}^\mr{id} - E_\mr{sol}^\mr{da}$ & \mbox{Difference in intraday and day-ahead solar energy forecast.} \\
		$E_\mr{won}^\mr{id}\!+\!E_\mr{woff}^\mr{id} - E_\mr{won}^\mr{da}\!-\!E_\mr{woff}^\mr{da}$ & \mbox{Difference in intraday and day-ahead wind energy forecast.} \\
		$E_\mr{sol}^\mr{id}\!\!+\!\!E_\mr{won}^\mr{id}\!\!+\!\!E_\mr{woff}^\mr{id} - E_\mr{sol}^\mr{da}\!\!-\!\!E_\mr{won}^\mr{da}\!\!-\!\!E_\mr{woff}^\mr{da}$ & \mbox{Difference in intraday and day-ahead renewables forecast.} \\\hline\hline %
		\multicolumn{2}{l}{Differences to time-lagged values of all features \phantom{\LARGE I}} \\\hline
		$\Delta_\mr{h}(x) = x_{(d,h)}(\tau) - x_{(d,h)-1\mr{h}}(\tau)$ & Maintaining forecast creation time $\tau$, the difference of all considered features $x_{(d,h)}$ for delivery hour $h$ on day $d$ to the value of that feature for the previous delivery hour is added to the feature space. \\
		$\Delta_\mr{d}(x) = x_{(d,h)}(\tau) - x_{(d,h)-1\mr{d}}(\tau\!-\!1\mr{d})$ & Changing the day of the forecast creation time $\tau$ to the previous day, the difference of all considered features $x_{(d,h)}$ for delivery hour $h$ on day $d$ to the value of that feature for the same delivery hour $h$ on the previous day $d-1$ is added to the feature space. \\\hline %
	\end{tabularx}
	\caption{Summary of all features calculated from base regressors listed in Table \ref{tab:regress_base} and external regressors given in Table \ref{tab:regress_base_ex}. The regressors from Tables \ref{tab:regress_base} and \ref{tab:regress_base_ex} combined with the variables defined here span the complete feature space for the forecast model.}
	\label{tab:regress_all}
\end{table*}

We summarise all additionally constructed features in Table \ref{tab:regress_all}, which, combined with the base regressors in Tables \ref{tab:regress_base} and \ref{tab:regress_base_ex}, represents the complete feature space fed into the feature selection procedure. The total feature space dimension amounts to $m=351$.

We organise the feature values in a $(n+1) \times m$ design matrix $X(d,h,\tau)$, i.e. each row constitutes a data point containing all $m$ feature values for a specific day $d-k$, where $d$ is the delivery day we are forecasting and $k=0,...,n$. For all data points, we fix the delivery hour $h$, as mentioned at the beginning of the section. All information that would only become available later than the forecast creation time $\tau$ is truncated, mainly pertaining live CID values and power forecasts. To ensure proper learning from the historical data, $\tau$ is set with respect to day $d$, and for each past day $d-k$ also shifted to this day, such that we get for $n+1$ data points
\begin{align}
	X(d,h,\tau) = \Big(& \;x\big(\,d\!-\!n,\, h,\, \tau\!-\!n\mr{d}\,\big),\; x\big(\,d\!-(\!n\!-\!1),\, h,\, \tau\!-(n\!-\!1)\mr{d}\,\big),\nonumber\\ &\; \dots,\;x\big(d,\, h,\, \tau\big) \;\Big)^\mr{T} , \label{eq:Xtau}
\end{align}
where $x(d-k,\, h,\, \tau)$ is a feature vector representing one data point pertaining delivery hour $h$ on delivery day $d-k$ with information available at creation time $\tau$. With $\tau-n\mathrm{d}$ we denoted shifting the creation time by $n$ days to the past, preserving the time information.

As target variable $\Y$, we use the end-of-day value of the IDFull index for day $d$, i.e. $\mr{eod}_d(P_\mr{idfull})$, which is available 5\,min before begin of delivery, i.e. $h-5\mr{min}$. With the target vector
\begin{align}
	\y(d,h) = \big(& \,\mr{eod}_{d-n}(P_\mr{idfull}),\;\mr{eod}_{d-(n-1)}(P_\mr{idfull}),\nonumber\\ &\; \dots,\; \mr{eod}_{d}(P_\mr{idfull}) \,\big)^\mr{T} , \label{eq:y}
\end{align}
we may formulate the regression model
\begin{align}
	y = w\,X + \epsilon, \label{eq:linreg}
\end{align}
with weight vector $w=(w_1,\dots,w_m)^\mr{T}$ and normal random variable $\epsilon\sim\NN(0,\sigma^2)^m$, omitting the dependence on $d$, $h$, $\tau$.

\subsection{Feature selection} \label{ssec:featsel}
In view of a total feature space dimension of $m=351$, and less than $n=730$ data points available per forecast, feature selection becomes mandatory. An additional difficulty arises from high collinearities between the features. Reasons for these collinearities include correlations between markets, shared causes for increasing and decreasing energy production and consumption, market participants acting on common ground, and features calculated from base regressors.

A common approach in the literature for automatic reduction of features is the least absolute shrinkage and selection operator (LASSO) \citep{tibshirani_regression_1996}. The LASSO is a $\ell_1$-regularisation technique, in which a penalty term
\begin{align}
	\|w\|_p=\bigg(\sum_{i=1}^m|w_i|^p\bigg)^{1/p}, \label{eq:lp}
\end{align}
is added to the least-square regression of \eqref{eq:linreg} with the choice $p=1$. The strength of the penalty is controlled via a Lagrange parameter $\lambda$. With increasing $\lambda$, more weights are shrunk to zero, thus performing effective feature selection. However, in case of strong collinearities, the LASSO tends to choose features that do not generalise well \citet{Su2017}.

Instead, we propose Orthogonal Matching Pursuit (OMP) \citep{pati1993orthogonal} for the task of feature selection. Matching Pursuit is a greedy algorithm that approximates the solution of the sparse regression problem corresponding to a $\ell_0$-regularisation via a stepwise iteration through feature space. The $\|w\|_0$ penalty imposes a constraint on the minimisation of the number of non-zero weights $w_i$ to the least-square regression, which has been shown to be superior over the LASSO and other regularisation techniques \citep{tropp_signal_2007,hastie_best_2020}, but belongs to the NP-hard complexity class. The orthogonal extension, OMP, additionally removes from the target variable the orthogonal projection of the feature selected to the target in each iteration step. This procedure improves convergence and provides additional robustness against collinearities. We used the algorithms implemented as \texttt{linear\_model.OrthogonalMatchingPursuit} and \texttt{linear\_model.LassoCV} in the \textsc{python} package \texttt{scikit-learn} \citep{rubinstein_efficient_2008,pedregosa_scikit-learn_2011}.

The hyperparameter for LASSO is the Lagrange parameter $\lambda$, while for OMP it is the maximum number of features, $n_\mr{feat}$. Typically, $\lambda$ for LASSO is set via cross-validation (CV). However, for better comparison with OMP, we explored optimising $\lambda$ to meet the $n_\text{feat}$ constraint but found that the CV built into \texttt{LassoCV} performed better. The hyperparameters of \texttt{LassoCV} had an insignificant influence on the results; hence, we used the default settings (e.g., 5-fold CV). For OMP, the greedy algorithm terminates when either the loss cannot be further improved or the cut-off $n_\mr{feat}$ is reached. We use $n_\mr{feat} = 20$, ensuring that the number of features determined by OMP typically stays well below the cut-off, except in a few cases with extreme outliers in the data.

To apply feature selection, the data is first cleaned: data points with a few missing values (e.g., occasional gaps in power forecasts, $<0.1\%$) are eliminated, and features with considerable missing values (e.g., live $P_\mr{id3}$ for $\tau < h-3$) are removed. Then, both the features in $X$ and the target vector $y$ are standardised to zero mean and unit variance. The selected features by OMP and LASSO are then taken from the original data, the cleaning procedure is repeated, and $X$ and $y$ are standardised again. In this way, most missing value problems resolve themselves after feature selection, reducing the need to remove missing data points to a minimum. Finally, zero-valued feature vectors (e.g., solar power at night) are removed if not already deselected by feature selection.

\section{The forecast model}\label{sec:fcmodel}
Based on the data model and feature selection described in the previous section, outputting design matrix \mbox{$X=X(d,h,\tau)$} and target vector \mbox{$y=y(d,h)$}, we now introduce the Bayesian forecast model used in this study. To this end, we separate historical data $(\Xtr,\ytr)$, comprising the $n$ data points for days $d\!-\!n,\dots,d\!-\!1$, from the most recent data point $(\Xpr,\ypr)$ for delivery day $d$. Note that at this stage $(\Xtr,\ytr)$ and $(\Xpr,\ypr)$ have already undergone feature selection.	

\begin{figure*}[t!]
	\begin{center}
		\includegraphics[width=0.45\textwidth]{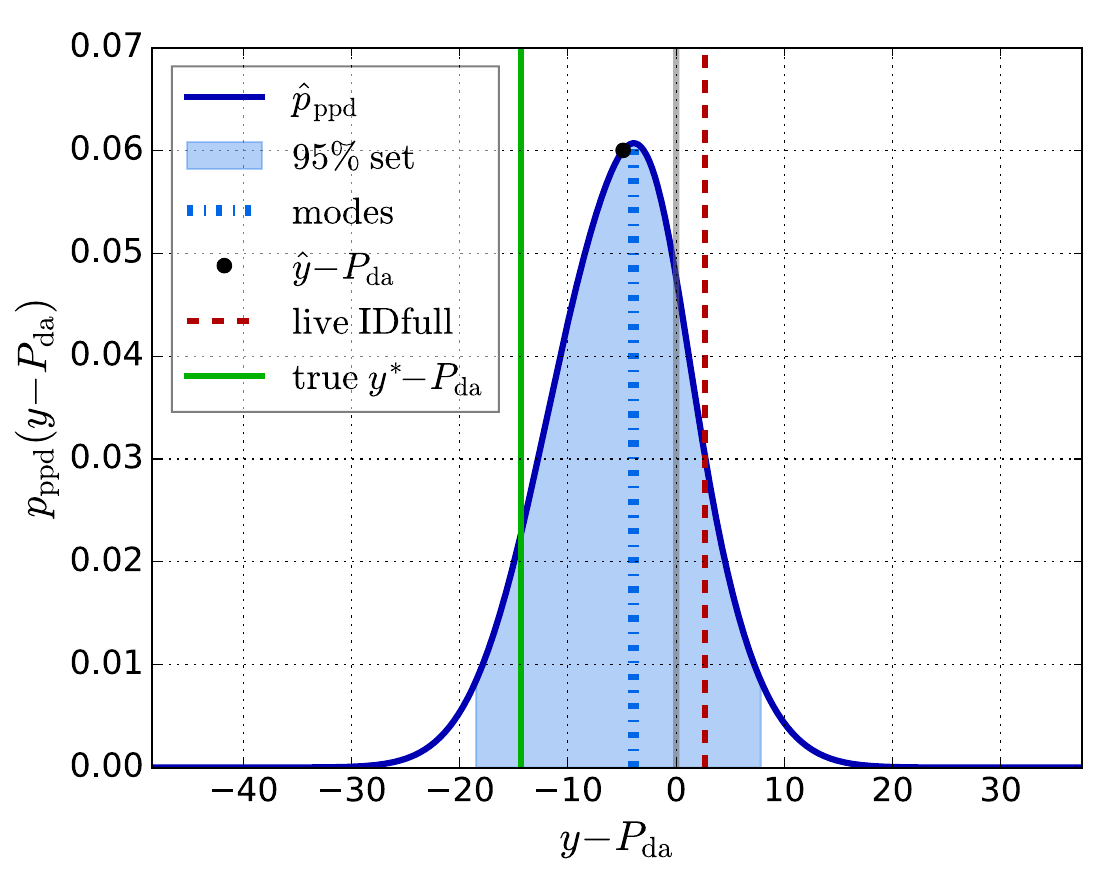} \hfil \includegraphics[width=0.45\textwidth]{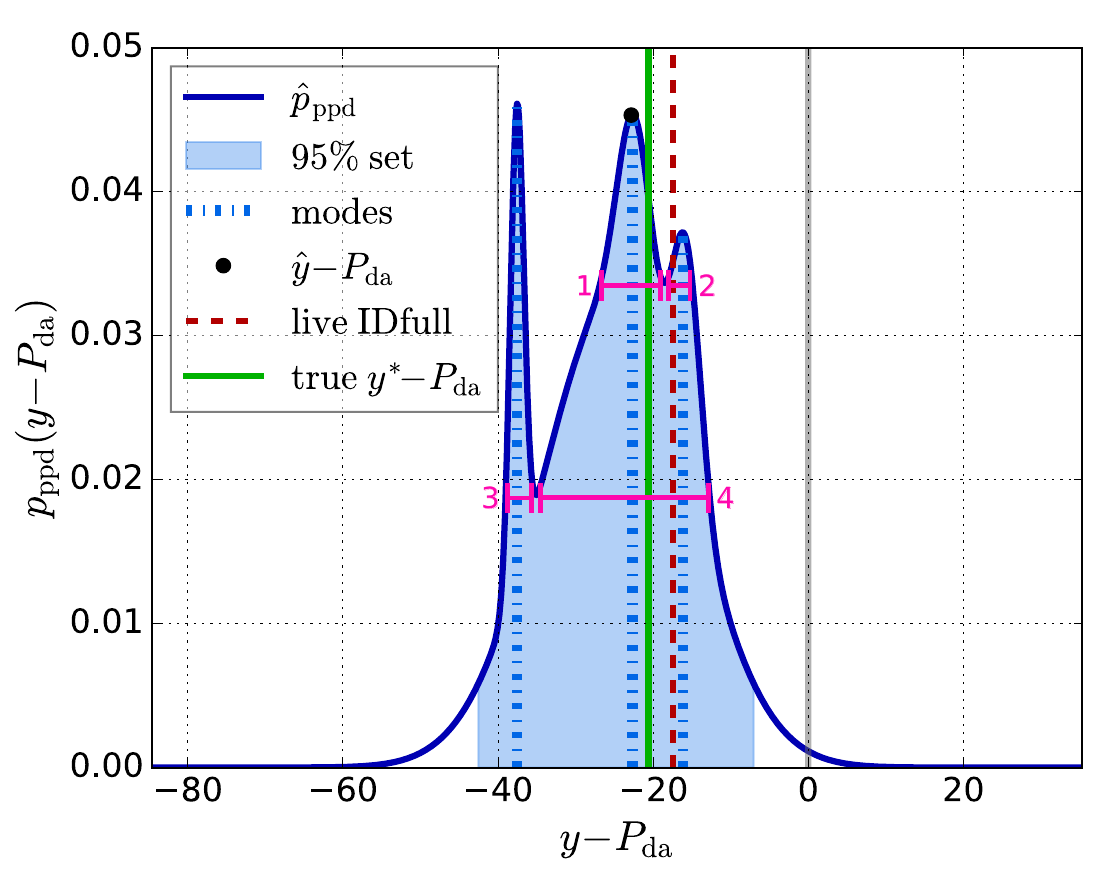}
	\end{center}
	\caption{A typical example of a forecast on the left (day $d=$ 24 August 2022, $h=0$) and a select example of a forecast on the right (day $d=$ 5 July 2022, hour $h=7$). The forecasts are represented by estimated posterior predictive distributions $\pde(y)$, from which credible intervals and point estimates are extracted. The example on the right, an extreme multimodal case showcasing the generality of the model, illustrates the procedure to pick a prediction interval and a point estimate using highest density intervals (HDIs). The intervals numbered 1, 2, 3, and 4 are formed by intersections with horizontal cuts, cf. \eqref{eq:hdi_def}. These intervals are special in that they would fuse into larger intervals for a slightly lower cut. From these fusing intervals, the interval with the largest credibility-to-width ratio is chosen, cf. \eqref{eq:PI_def}, which in this example is interval 1. This interval serves as the PI $\hat I$, and the median within this interval as the point estimate $\hat y$, cf. \eqref{eq:yhat_def}.}
	\label{fig:ex_ppd}
\end{figure*}

As a probabilistic model for the random target variable $Y$, we extract a Gaussian likelihood from the linear model \eqref{eq:linreg},
\begin{align}
	\lk(\y\,|\,\Xtr,w,\sg) = \NN(w\Xtr,\sigma^2) . \label{eq:like}
\end{align}
Note that any distribution could be chosen here, but without indication of a specific uncertainty structure, we apply the principle that the simplest explanation is usually the best. A Gaussian likelihood also fits well with the empirical Bayes approach explained below. The generality of the model lies in the fact that the posterior predictive distribution is a compound distribution between the likelihood and the posterior, which can take a very general form, as illustrated in Figure \ref{fig:ex_ppd} further below.

For the weight vector $w$ we choose the product of normal distributions as prior and independently the Gamma distribution for the standard deviation $\sigma$,
\begin{align}
	\pr(w,\sg) = \NN(\mu_w, \sg_w) \cdot \GG(\a,\b) . \label{eq:prior} 
\end{align}
This choice of prior imposes $\ell_2$-regularisation on $w$ (Ridge regression), cf. \eqref{eq:lp}. We also tested a Laplacian prior (equivalent to $\ell_1$-regularisation) for $w$ but did not find any notable improvement. The conjugate prior for $w$ and $\sg^2$ would be the (multivariate) normal-inverse-gamma distribution, but we prefer to set the priors for $w$ and $\sg$ independently. This approach facilitates the empirical Bayes method used below to reduce the number of hyperparameters and has the additional benefit of not fixing the form of the posterior to the conjugate family, thus adding to the generality of the model.

The parameters $\mu_w$ and $\sg_w$ of the prior for weights $w$ are determined in an empirical Bayes approach. For the mean we choose the ordinary least-square (OLS) estimate $\mu_w=C\,\Xtr^\mr{T}\,\ytr$ with $C=(\Xtr^\mr{T}\,\Xtr)^{-1}$. For the standard deviation we make use of the normality of OLS estimates and set $\sg_w = \frac{1}{n}\,S\,\mr{diag}(C)$ with $S = (\ytr - \mu_w\,\Xtr^\mr{T})\,\ytr$. Similarly, we set the mode $m_\sg=\frac{\a-1}{\b}$ of the Gamma distribution to $1$ to account for standardised data, i.e. $\a=m_\sg\b+1$. Finally, to still ensure weakly informative priors, we found that $\b=\frac{1}{2}$ is an adequate choice, which corresponds to a variance of $\frac{\alpha}{\beta^2}=6$. Other moderate choices for the only remaining hyperparameter $\beta$ turned out to not have a significant influence on the results.

\begin{figure*}[t!]
	\begin{center}
		\includegraphics[width=0.95\textwidth]{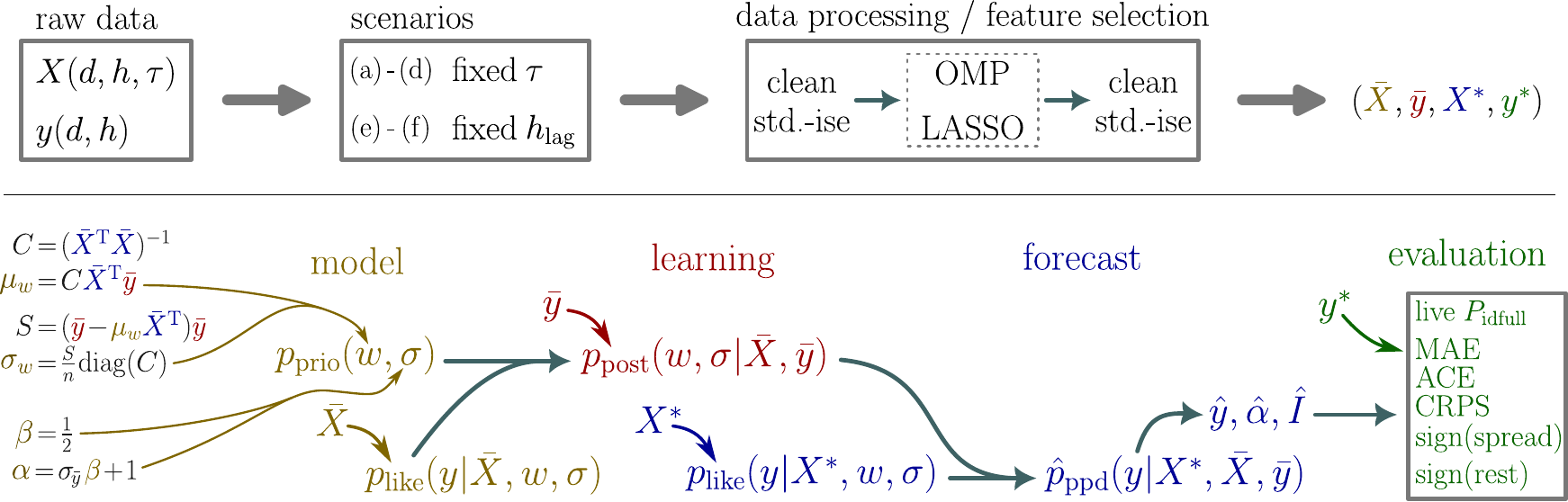}
	\end{center}
	\caption{A schematic overview of the data model (top row) and the forecast model (bottom row). The data model takes the design matrix $X(d,h,\tau)$ and target values $y(d,h)$ as input, cf. \eqref{eq:Xtau} and \eqref{eq:y}. This input data is then organised according to forecast scenarios described in Section \ref{sec:results}, where either the forecast creation time $\tau$ or the lag $\hlag$ between forecast creation and delivery begin is fixed. Subsequently, the data is cleaned for missing values, standardised and undergoes feature selection (OMP for scenarios (a)-(e) and LASSO for (f)), as described in Section \ref{ssec:featsel}. The historical data $(\Xtr,\ytr)$ enters the forecast model for learning, the new data point $(\Xpr,\ypr)$ is used for forecast creation and evaluation. The hyperparameters $\mu_w$, $\sigma_w$, $\alpha$, $\beta$ of the prior $\pr$ are defined in an empirical Bayes approach using $(\Xtr,\ytr)$, as explained in Section \ref{sec:fcmodel}. The posterior $\po$ follows with Bayes formula from prior $\pr$ and likelihood $\pl$, and the estimated posterior predictive distribution $\pde$ as the compound distribution of $\po$ and $\pl$ at the new datapoint $\Xpr$, both explicated in Section \ref{sec:fcmodel}. The procedure to extract point estimates $\hat y$ and prediction intervals $\hat I$ at credibility levels $\hat\alpha$ from $\pde$ is explained in Section \ref{ssec:probfor}, and their evaluation with respect to the true value $\ypr$ and the live IDFull as benchmark is given in Section \ref{ssec:eval} in terms of mean absolute error (MAE), average coverage error (ACE), continuous ranked probability score (CRPS), and signs of \textit{spread} and \textit{rest} values.}
	\label{fig:summary}
\end{figure*}

With Bayes' theorem, we can write down the posterior distribution as
\begin{align}
	\po(w,\sg\,|\,\Xtr,\ytr) = \frac{\lk(\Y=\ytr\,|\,\Xtr,w,\sg) \cdot \pr(w,\sg)}{\pv(\ytr\,|\,\Xtr)} , \label{eq:post}
\end{align}
where we have in the denominator the prior-predictive value
\begin{align}
	\pv(\ytr\,|\,\Xtr) = \int_0^\infty \int_{\RR^m} &\,\lk(\Y=\ytr\,|\,\Xtr,w,\sg) \nonumber\\&\quad\cdot \pr(w,\sg) \;\mr{d}^mw\,\mr{d}\sigma . \label{eq:ppv}
\end{align}
Note that, due to its independence from $w$ and $\sigma$, $\pv(\ytr\,|\,\Xtr)$ can be ignored in sampling or maximisation of the posterior $\po(w,\sg\,|\,\Xtr,\ytr)$. 

Bayesian point estimates for $w$ and $\sigma$ may now be obtained by maximizing the posterior $\po(w,\sg\,|\,\Xtr,\ytr)$ (MAP estimates), which plugged into the likelihood \eqref{eq:like} and replacing $\Xtr$ by $\Xpr$ yields a probabilistic forecast for the target $\Y$. However, in doing so, we would discard valuable information about the uncertainty of $w$ and $\sigma$ captured by the posterior distribution. To fully incorporate this uncertainty, we instead determine the posterior predictive distribution (PPD), 
\begin{align}
	\pd(\y\,|\, \Xpr,\Xtr,\ytr) = \int_0^\infty \int_{\RR^m} &\,\lk(\y\,|\,\Xpr,w,\sg) \nonumber\\&\;\cdot \po(w,\sg\,|\,\Xtr,\ytr) \;\mr{d}^mw\,\mr{d}\sigma . \label{eq:ppd}
\end{align}
Note the different roles played by $(\Xtr,\ytr)$ and $\Xpr$. The historical data is used to set up the posterior by virtue of simple substitution instead of training as in classical machine learning approaches. The new data point $\Xpr$ is then used in the likelihood to produce the forecast distribution $\pd(\y\,|\, \Xpr,\Xtr,\ytr)$ for the end-of-day IDFull.

However, instead of computational intensive training, we here face a sampling problem to evaluate the high-dimensional integral. Setting up the posterior distribution in the \textsc{python} package \texttt{TensorFlow Probability} (TFP) \citep{dillon_tensorflow_2017} in the acceleration environment \texttt{JAX} \citep{bradbury_jax_2018}, we can leverage efficient implementations of Hamilton Monte-Carlo (HMC). The best performance was obtained by the No-U-Turn Sampler (NUTS) \citep{Hoffman2014no}, a recursive adaptation of HMC that automatically detects the reversion to already sampled parts of parameter space, thus ensuring a more thorough sampling and disposes of the difficult choice of the number of iteration steps. The parameters of the NUTS algorithm built into TFP are set to an initial step size of $0.001$ and a burn-in period of $500$ iterations, both of which were found to be a good universal compromise between accuracy and computational efficiency in our application.

On a standard laptop, $N=140'000$ posterior samples $w_i$ and $\sigma_i$ of the parameter vector $(w,\sigma)^\mr{T}$ were thus obtained in matters of seconds for each forecast. With these samples, we estimate the PPD by substituting the posterior expectation with an average,
\begin{align}
	\pde(\y\,|\, \Xpr,\Xtr,\ytr) = \frac{1}{N} \sum_{i=1}^N &\,\lk(\y\,|\,\Xpr,w_i,\sg_i) . \label{eq:ppd_esti}
\end{align}

The data model and the forecast model are summarised in Figure \ref{fig:summary}.

\section{Forecast study and discussion}\label{sec:results}

To test our model, we used data from 2021 and 2022 to build $X$ and $y$ according to \eqref{eq:Xtau} and \eqref{eq:y} and Tables \ref{tab:regress_base}-\ref{tab:regress_all}. We consider 6 different forecast scenarios distinguished by forecast creation time $\tau$. In a first set of 4 scenarios, we fix $\tau$ and forecast the end-of-day IDFull $\Y=\mr{eod}_d(P_\mr{idfull})$ for all hours $h>\tau$ of delivery day $d$, 
\begin{itemize}
	\item[(a)] $\boldsymbol{\tau=\;}$\textbf{23:00\,pm on day} $\boldsymbol{d\!-\!1}$, \newline delivery hours $\boldsymbol{h=0,...,23}$ on day $d$ forecasted,
	\item[(b)] $\boldsymbol{\tau=\;}$\textbf{5:00\,am on day} $\boldsymbol{d}$, \newline delivery hours $\boldsymbol{h=6,...,23}$ on day $d$ forecasted,
	\item[(c)] $\boldsymbol{\tau=\;}$\textbf{11:00\,am on day} $\boldsymbol{d}$, \newline delivery hours $\boldsymbol{h=12,...,23}$ on day $d$ forecasted,
	\item[(d)] $\boldsymbol{\tau=\;}$\textbf{17:00\,pm on day} $\boldsymbol{d}$, \newline delivery hours $\boldsymbol{h=18,...,23}$ on day $d$ forecasted.
\end{itemize}
In addition to these four scenarios, where the forecast creation time is fixed, we also consider a scenario in which we fix the lag $\hlag=h-\tau$ between forecast creation and begin of delivery. For this scenario, we always forecast all hours of the delivery day, and consider lags of $\hlag=1,...,6$. We use this more extensive scenario to compare the forecast performance when OMP is used for feature selection with the case where LASSO is used for feature selection. Thus we have the two additional cases	
\begin{itemize}
	\item[(e)] $\tau=h-\hlag$, hours $h=0,...,23$ forecasted on day $d$, \newline for $h_\mathrm{lag}=1,...,6$, using \textbf{OMP},
	\item[(f)] $\tau=h-\hlag$, hours $h=0,...,23$ forecasted on day $d$, \newline for $h_\mathrm{lag}=1,...,6$, using \textbf{LASSO}.
\end{itemize}

To evaluate the forecast performances of these scenarios, we consider $183$ test days $d$ covering the second half of 2022 (1 July - 30 December). For each test day $d$, all hours $h$ are forecasted as specified in the scenarios above. Each forecast uses all previous days $d-1,d-2,...,d-n$ to construct $(\Xtr,\ytr)$ for learning the posterior \eqref{eq:post}, and $\Xpr$ of delivery day $d$ is used to estimate the forecast distribution \eqref{eq:ppd_esti}, $\pde(\y\,|\, \Xpr,\Xtr,\ytr)$. For the scenarios (e) or (f), for instance, the forecast study consists of $183\times24=4392$ individual forecasts for each of the $6$ values for $\hlag$.

\begin{figure*}[t!]
	\phantom{X}{Scenario (a), 6 July 2022} \hfil\phantom{XX} {Scenario (e), 5 October 2022} \hfil\\[-7mm]
	\begin{center}
		\includegraphics[width=0.49\textwidth]{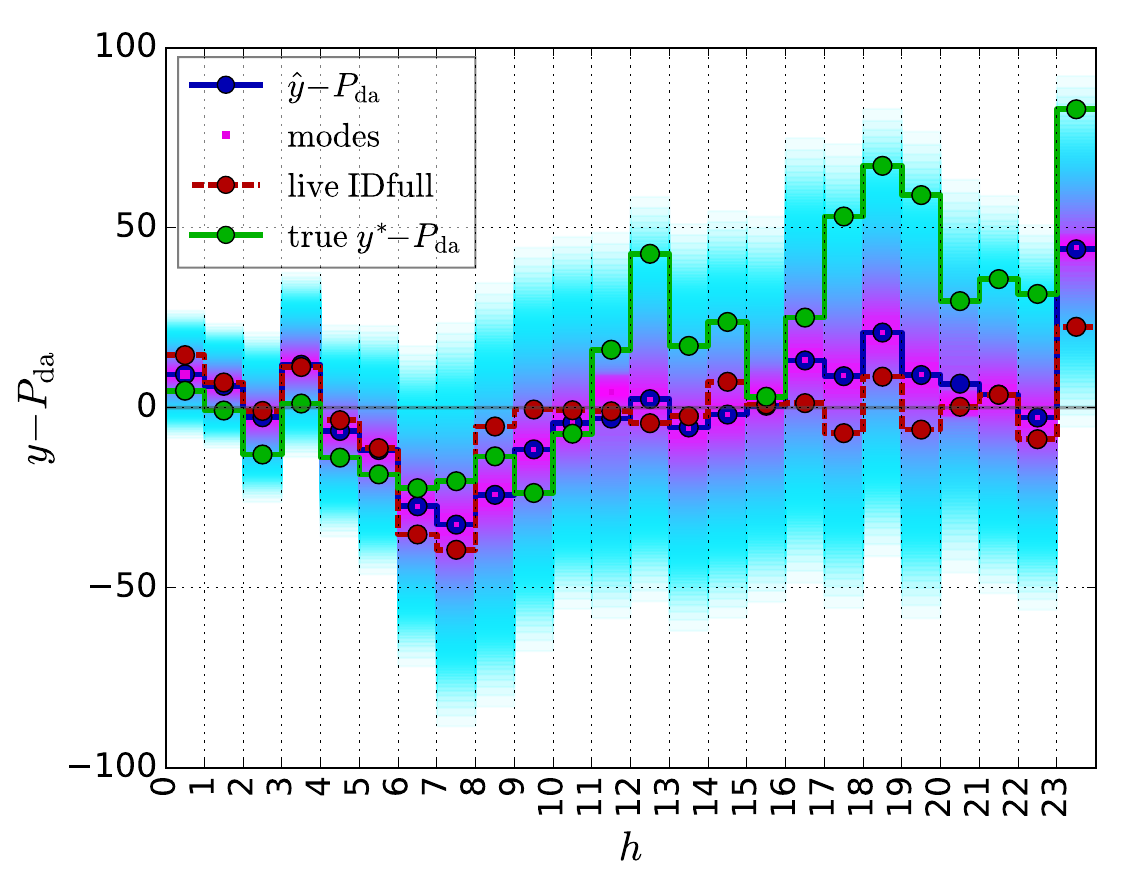} \hfil \includegraphics[width=0.49\textwidth]{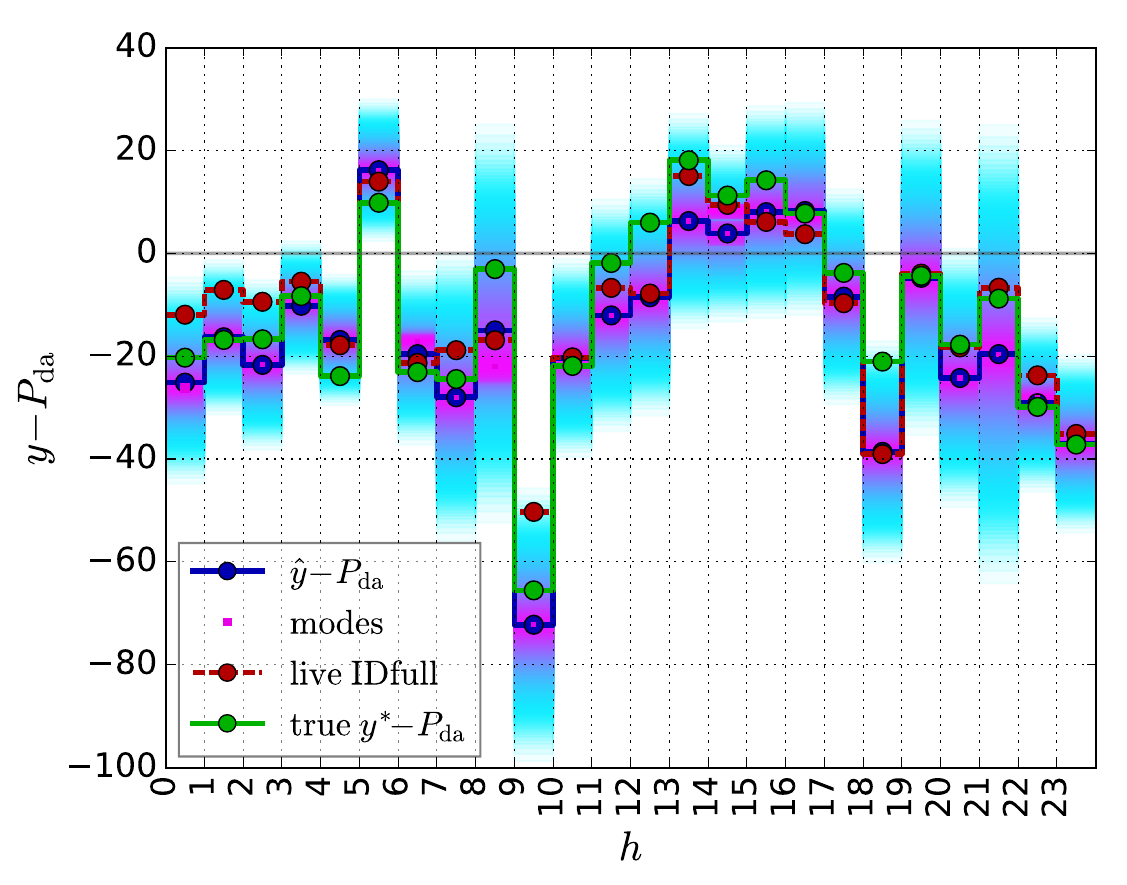}
	\end{center}
	\caption{Exemplary full day forecasts for all delivery hours $h$ of day $d=\;$6 July 2022 on the left and all delivery hours $h$ of day $d=\;$5 October 2022 on the right. The forecast on the left is part of scenario (a), where the forecast creation time $\tau$ is fixed to 23:00\,pm on day $d-1$, implying an increasing forecast lag of $\hlag=h+1$ along the horizontal axis. The forecast on the right is part of scenario (e), where the lag between forecast and delivery begin is fixed to 1 hour, i.e. $\hlag=1$, implying forecast creation times $\tau=h-1$. The vertical colour bars represent the topology of the respective predictive distributions $\pde(y)$ by means of $\alpha(p_\mr{cut})$ for 100 values of $p_\mr{cut}$, cf. \eqref{eq:hdi_def}.}
	\label{fig:ex_ppd_day}
\end{figure*}

\subsection{Probabilistic forecasts} \label{ssec:probfor}
We illustrate our forecast procedure with a few examples. Two individual forecasts are shown in Figure \ref{fig:ex_ppd}. Since a typical reference price for the CID market is the DA price $P_\mr{da}$, we consider the spread between the IDFull and the DA price, $\Y-P_\mr{da}$. Due to the probabilistic forecast, there are several options to extract a point estimate from the estimated predictive distribution $\pde(\y)$, e.g. the mode, median or mean of $\pde(\y)$. In terms of modes, an additional ambiguity arises when $\pde(\y)$ is multimodal. Once a point estimate is extracted from $\pde(\y)$, credible intervals of various credibility levels $\alpha$ can be extracted in various ways, which shall serve as prediction intervals (PIs) in this study. We found the following procedure to be best in terms of robustness and accuracy in our forecast study.

Most prominent examples of determining credible intervals from predictive distributions are percentile intervals or highest density intervals (HDIs) \citep{mcelreath_statistical_2020}. Here we choose HDIs as they are more suitable for possibly skewed distributions and also naturally handle multimodal distributions \citep{hyndman_computing_1996,hyndman_estimating_1996}. Specifically, to determine $\alpha$-credible HDIs, we determine a value $p_\mr{cut}$ such that intersections ${l_1,u_1,\dots,l_k,u_k}$ of $\pde(y)$ with $p_\mr{cut}$ determine a set $[l_1,u_1]\cup\dots\cup[l_k,u_k]$ that cover a fraction $\alpha$ of the total probability mass,
\begin{align}
	\alpha(p_\mr{cut}) = \sum_{i=1}^k \int_{l_i}^{u_i} \pde(y) \,\mr{d}y ,   \label{eq:hdi_def}
\end{align}
with $\pde(l_i)=\pde(u_i)=p_\mr{cut} \;\forall i\in\{1,...,k\}$ .
Conducting this procedure for an decreasing list of cut values $p_\mr{cut}$, we obtain a list of credible sets $[l_1(p_\mr{cut}),u_1(p_\mr{cut})]\cup\dots\cup[l_k(p_\mr{cut}),u_k(p_\mr{cut})]$ with increasing credibilities $\alpha(p_\mr{cut})$. Typically, for high values of $p_\mr{cut}$, single credible intervals (i.e. $k=1$) are determined by the highest peak of $\pde$, whereas for $p_\mr{cut}\to0$ the domain of $\pde$ becomes the credible interval for $\alpha(0)=1$. 

For intermediate $p_\mr{cut}$, sub-intervals may develop which fuse again to larger credible intervals as $p_\mr{cut}$ is decreased. In order to choose a PI from all intervals $[l_i(p_\mr{cut}),u_i(p_\mr{cut})]$, we consider all fusing intervals $\{I_\mr{fuse}\}$, and of these pick $\hat I=[\hat y_l,\hat y_u]$ that maximizes the credibility-to-width ratio,
\begin{align}
	\hat I = \argmax_{I\in \{I_\mr{fuse}\}} \frac{1}{|I|} \int_{I} \pde(y) \,\mr{d}y , \label{eq:PI_def}
\end{align}
accompanied by a credibility of $\hat\alpha=\int_{\hat I} \pde(y) \,\mr{d}y$.
As a point estimate $\hat y$, we derive the median within this interval, that is, $\hat y$ is chosen such that
\begin{align}
	\int_{\hat y_l}^{\hat y} \pde(y) \,\mr{d}y = \frac{\hat\alpha}{2}. \label{eq:yhat_def}
\end{align}

This procedure to determine PIs $\hat I$ and point estimates $\hat y$ in the general case of multimodal predictive distributions is also illustrated in Figure \ref{fig:ex_ppd}. In Figure \ref{fig:ex_ppd_day}, we show two examples in which the probabilistic forecasts for all hours of a day are depicted. These full day forecasts demonstrate the typical behaviour of our forecast model, namely identifying the live IDFull as the most important regressor, supporting the weak-form efficiency hypothesis. Nevertheless, the point estimates $\hat y$ often do correct the live IDFull values towards the ground truth $\ypr$. We furthermore observe that forecasts with a larger lag between creation time $\tau$ and delivery begin $h$ are characterised by broader prediction intervals, indicating that our forecast model correctly represents forecast uncertainty.

In the following evaluation of our forecast study, we quantify and assess the significance of these observation, where the live IDFull $P_\mr{idfull}$ shall serve as a reference.

\subsection{Evaluation of point estimates}\label{ssec:eval}
\begin{figure*}[t!]
	\phantom{X}{Scenario (a), $\tau=23$, $d-1$} \hfil \phantom{X}{Scenario (b), $\tau=5$, $d$}   \hfil\\[-7mm]
	\begin{center}
	\includegraphics[width=0.45\textwidth]{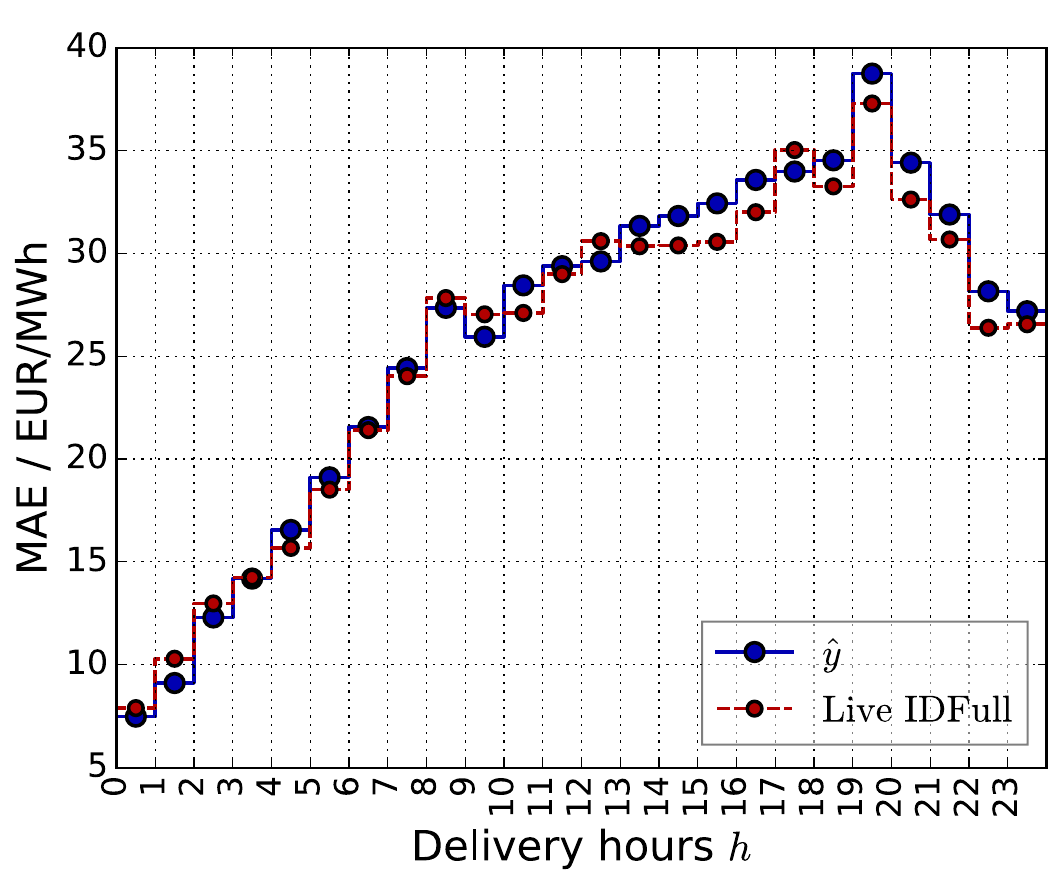} \hfil \includegraphics[width=0.45\textwidth]{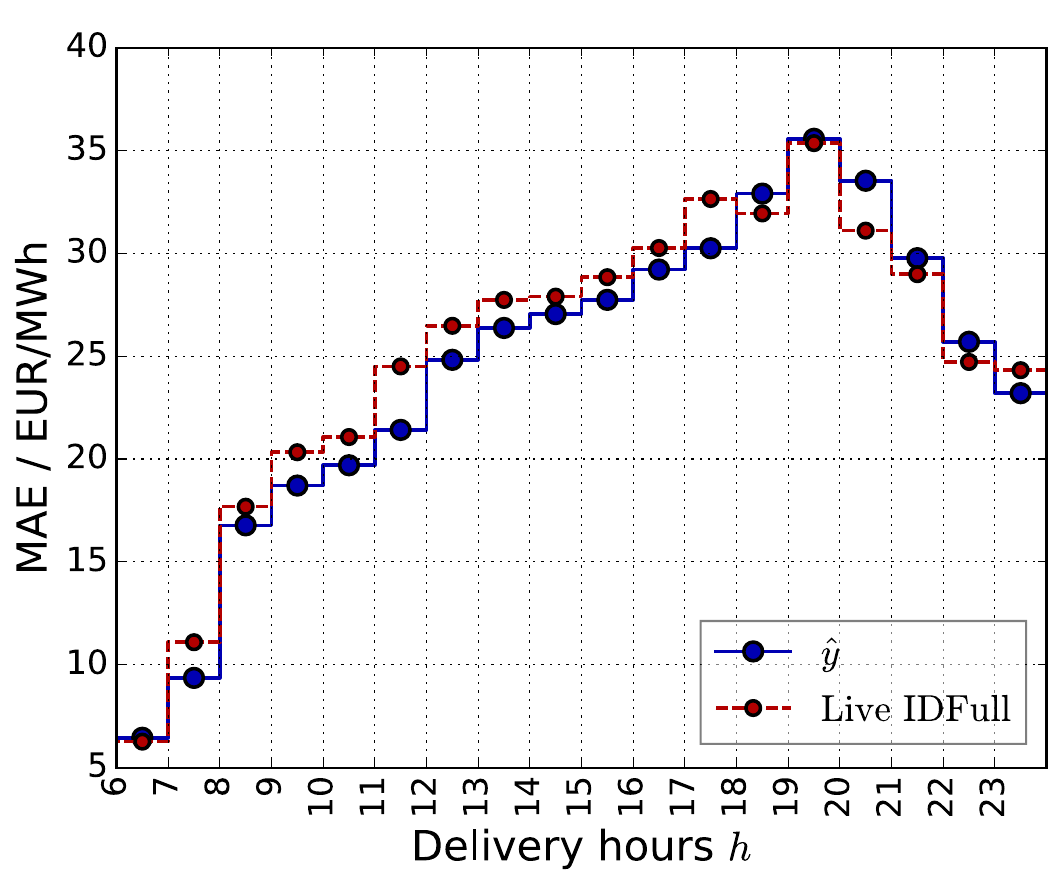}
	\end{center}
	\phantom{X}{Scenario (c), $\tau=11$, $d$} \hfil \phantom{X}{Scenario (d), $\tau=17$, $d$} \hfil\\[-7mm]
	\begin{center}
	\includegraphics[width=0.45\textwidth]{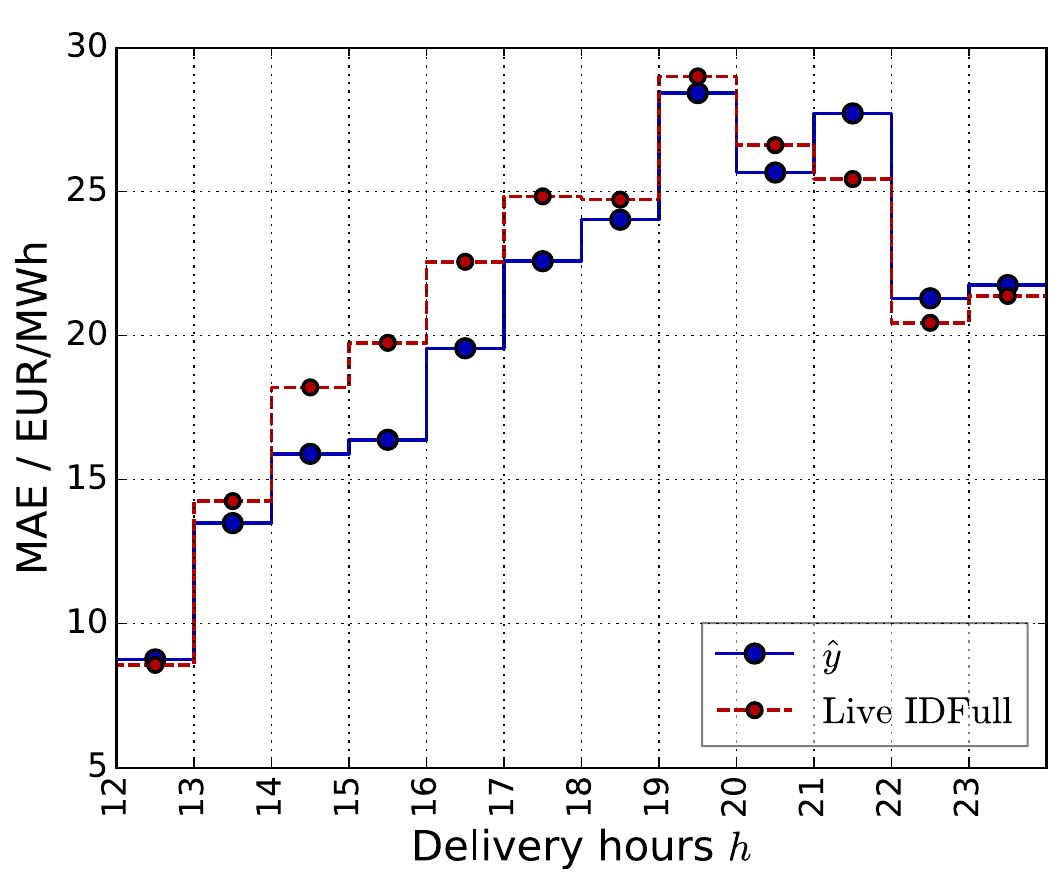} \hfil \includegraphics[width=0.45\textwidth]{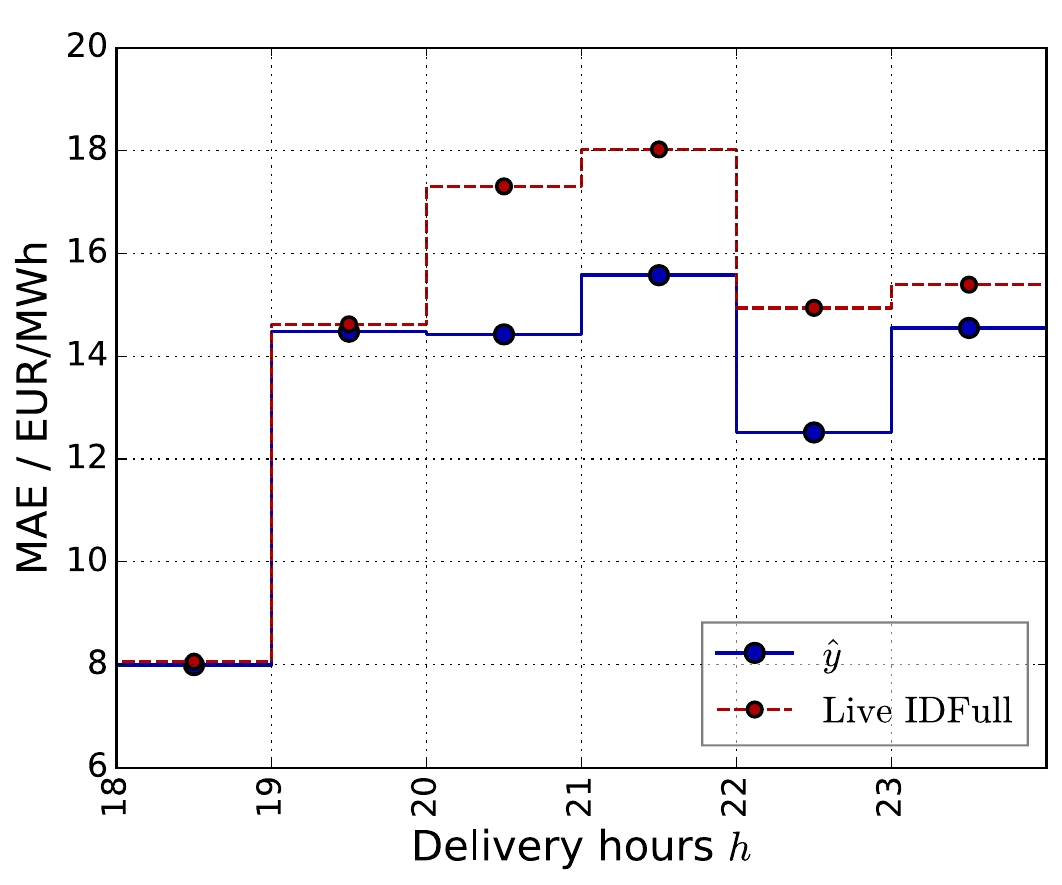}
	\end{center}
	\caption{Mean absolute errors (MAEs) for scenarios (a) - (d), where forecast creation time is fixed, and all following delivery hours are forecasted. The point estimate $\hat\y$ for the end-of-day IDFull is given by \eqref{eq:yhat_def}, the live IDFull $P_\mr{idfull}$ is used as reference.}
	\label{fig:MAE_abcd}
\end{figure*}

As a first evaluation, we compute the mean absolute error (MAE) of point estimates $\hat y$ for all scenarios (a) - (f). As a benchmark model, we use the live IDFull $P_\mr{idfull}$, in line with the proposed weak-form efficiency assumption of the CID market. In Figure \ref{fig:MAE_abcd}, we show the MAE of $\hat y$ and the live IDFull for all forecast hours of scenarios (a) - (d). These results indicate that with more transaction data available, not only the MAE becomes smaller, also the forecast model tends to beat the live IDFull more often.

\begin{figure*}[t!]
	\phantom{X}{Scenarios (a) - (d)} \hfil \phantom{X}{ Scenarios (e) and (f)} \hfil\\[-7mm]
	\begin{center}
	\includegraphics[width=0.45\textwidth]{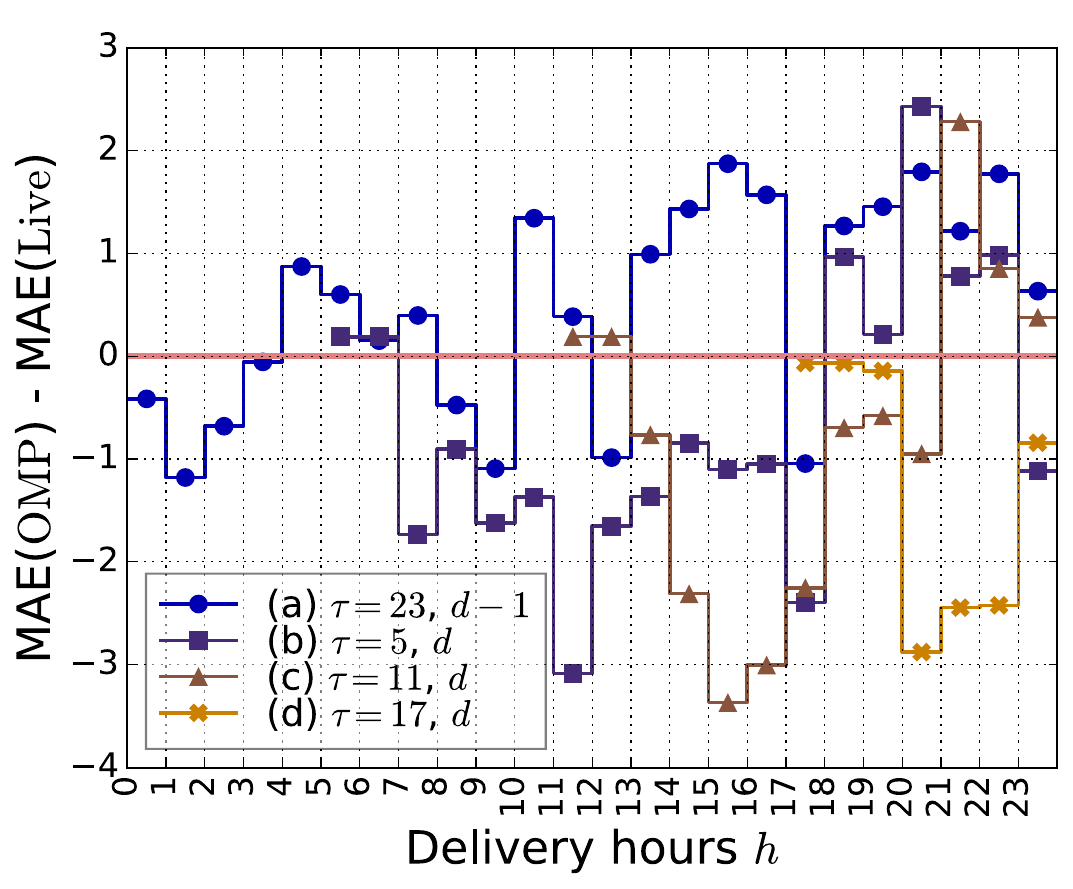} \hfil \includegraphics[width=0.45\textwidth]{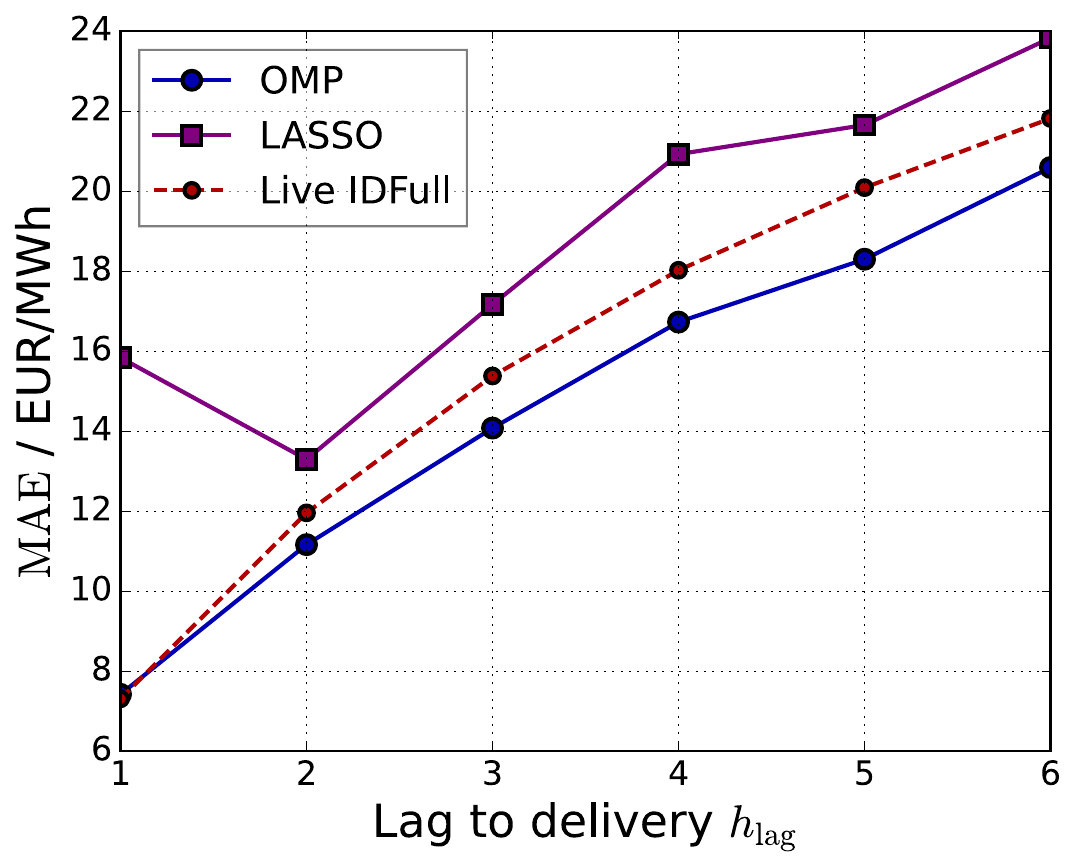}
	\end{center}
	\phantom{X}{Scenario (e) \;(OMP feature selection)} \hfil \phantom{X}{Scenario (f) \;(LASSO feature selection)} \hfil\\[-7mm]
	\begin{center}
	\includegraphics[width=0.45\textwidth]{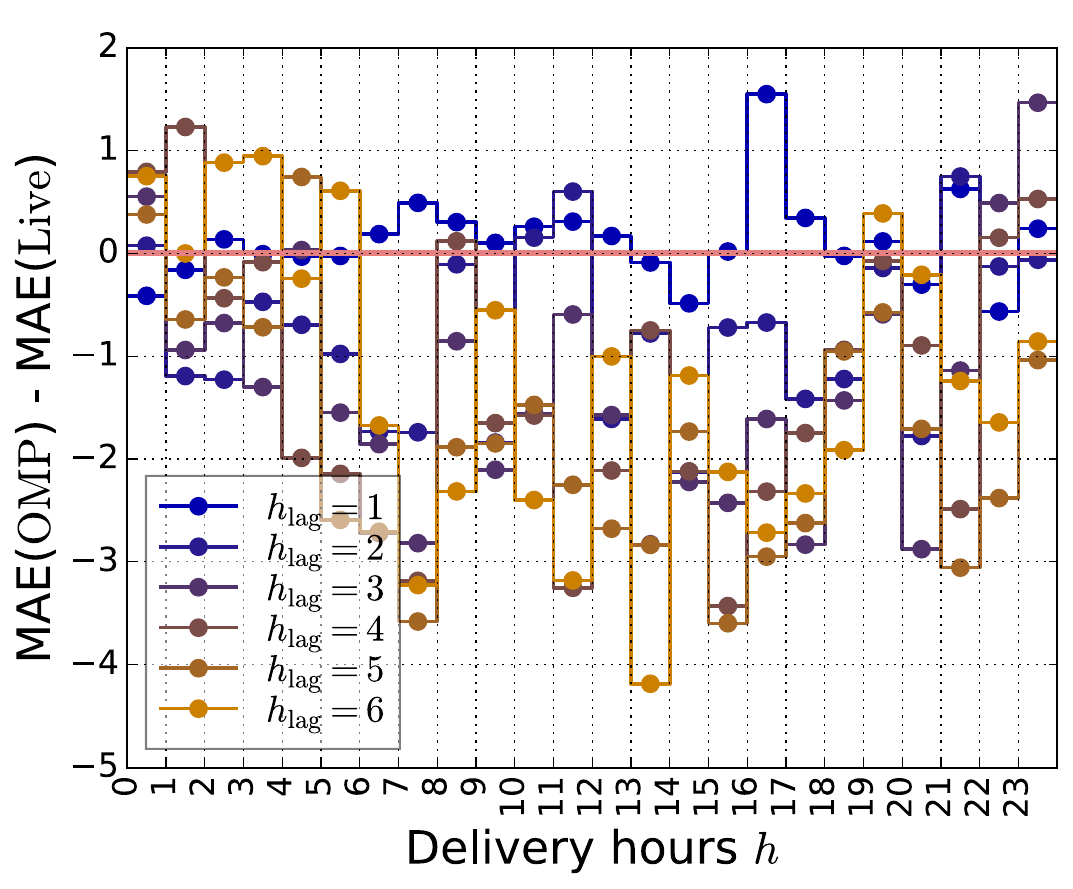} \hfil \includegraphics[width=0.45\textwidth]{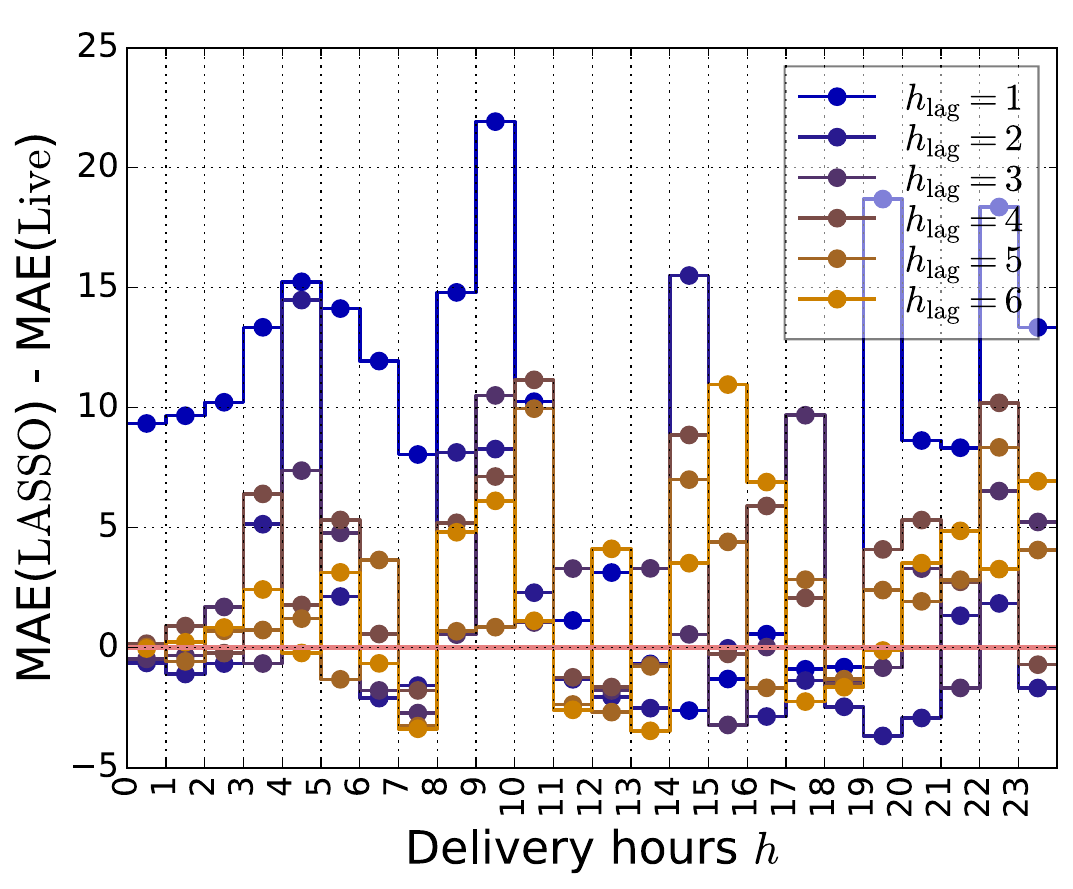}
	\end{center}
	\caption{The top left chart depicts the difference between $\hat\y$ and the live IDFull in terms of their MAE for scenarios (a) - (d), summarizing Figure \ref{fig:MAE_abcd}. The bottom shows the same for scenarios (e) using OMP and (f) using LASSO for feature selection. Results below the solid red line possess smaller MAEs than their live IDFull references. In the top right chart, the MAE values are averaged across hours, with OMP showing an average decrease of $5.9\,\%$ in absolute errors compared to the live benchmark and a $22.7\,\%$ reduction compared to LASSO, while LASSO leads to an $28.8\,\%$ increase of absolute errors compared to the live benchmark.}
	\label{fig:MAE_abcdef}
\end{figure*}

A more condensed form, now also including scenarios (e) and (f), confirms this observation in Figure \ref{fig:MAE_abcdef}, where the difference of the MAE between $\hat y$ and the live IDFull is shown, as well as the MAE averaged across all days and hours for scenarios (e) and (f). We observe that $\hat y$ in scenario (e) beats the live IDFull for all $\hlag\geq2$, and leads to considerable smaller MAE compared to scenario (f) for all $\hlag$. The fact that scenario (e) achieves the overall smallest forecast errors challenges the weak-form efficiency hypothesis and implies that OMP feature selection is superior over LASSO. Later we show that these results are statistically significant.

\subsection{Spread sign forecasts}
Given the probabilistic forecast, more than point estimates can be extracted. A straight forward extraction are the probabilities to observe an IDFull smaller or larger than the DA price, that is the sign of the \textit{spread} $\Delta P_\mr{spread} = \mr{eod_d}(P_\mr{idfull})-P_\mr{da}$, which can be estimated as
\begin{align}
	\hat p_-^\mr{spread} &= \int_{-\infty}^{P_\mr{da}} \pde(\y) \,\d\y \,, \quad \hat p_+^\mr{spread} = \int_{P_\mr{da}}^{\infty} \pde(\y) \,\d\y . \label{eq:p-p+_spread}
\end{align}
The sign of the spread is of practical relevance, as it informs a seller on the CID market when to sell electricity for a higher price than the DA price. The probabilities $\hat p_-^\mr{spread}$ and $\hat p_+^\mr{spread}$ can be used to maximize profit and minimize risk \citep{Maciejowska2019,Maciejowska_enhancing_2021}.

Also of interest is the \textit{rest} $\Delta P_\mr{rest} =\mr{eod_d}(P_\mr{idfull})-P_\mr{idfull}$,  that is the difference of the end-of-day value of the IDFull to its current (live) value. A positive $\Delta P_\mr{rest}$ indicates an average increase of electricity prices until gate closure, and a negative sign indicates falling prices. The probabilities of these two cases, estimated by
\begin{align}
	\hat p_-^\mr{rest} &= \int_{-\infty}^{P_\mr{idfull}} \pde(\y) \,\d\y \,, \quad \hat p_+^\mr{rest} = \int_{P_\mr{idfull}}^{\infty} \pde(\y) \,\d\y , \label{eq:p-p+_rest}
\end{align}
are therefore again of practical relevance.

To evaluate these probabilistic sign forecasts, we consider the following estimator for the sign $s$ of $\Delta P_\mr{spread}$, based on a credibility threshold $p_0\geq0.5$,
\begin{align}
	\hat s(\mr{spread}) = 
	\begin{cases}
	+1 & p_+^\mr{spread} > p_0 \,, \\
	\mr{sgn}(P_\mr{idfull}) & 1-p_0 \leq p_+^\mr{spread} \leq p_0 \,, \\
	-1 & p_-^\mr{spread} > p_0 \,,
	\end{cases} \label{eq:sgn_spread}
\end{align}
that is, if we exceed the credibility threshold, we use the forecasted sign implied by $\pde(y)$, otherwise we use the sign of the live IDFull. For the sign of $\Delta P_\mr{rest}$, we use the simpler estimator
\begin{align}
	\hat s(\mr{rest}) = 
	\begin{cases}
	+1 & p_+^\mr{rest} > 0.5 \,, \\
	-1 & p_-^\mr{rest} > 0.5 \,,
	\end{cases} \label{eq:sgn_rest}
\end{align}
since $p_\pm^\mr{rest}$ already involves $P_\mr{idfull}$.

\begin{figure*}[t!]
	\phantom{X}{Scenario (a), $\tau=23$, $d-1$} \hfil \phantom{X}{ Scenario (d), $\tau=17$, $d$} \hfil\\[-7mm]
	\begin{center}
	\includegraphics[width=0.45\textwidth]{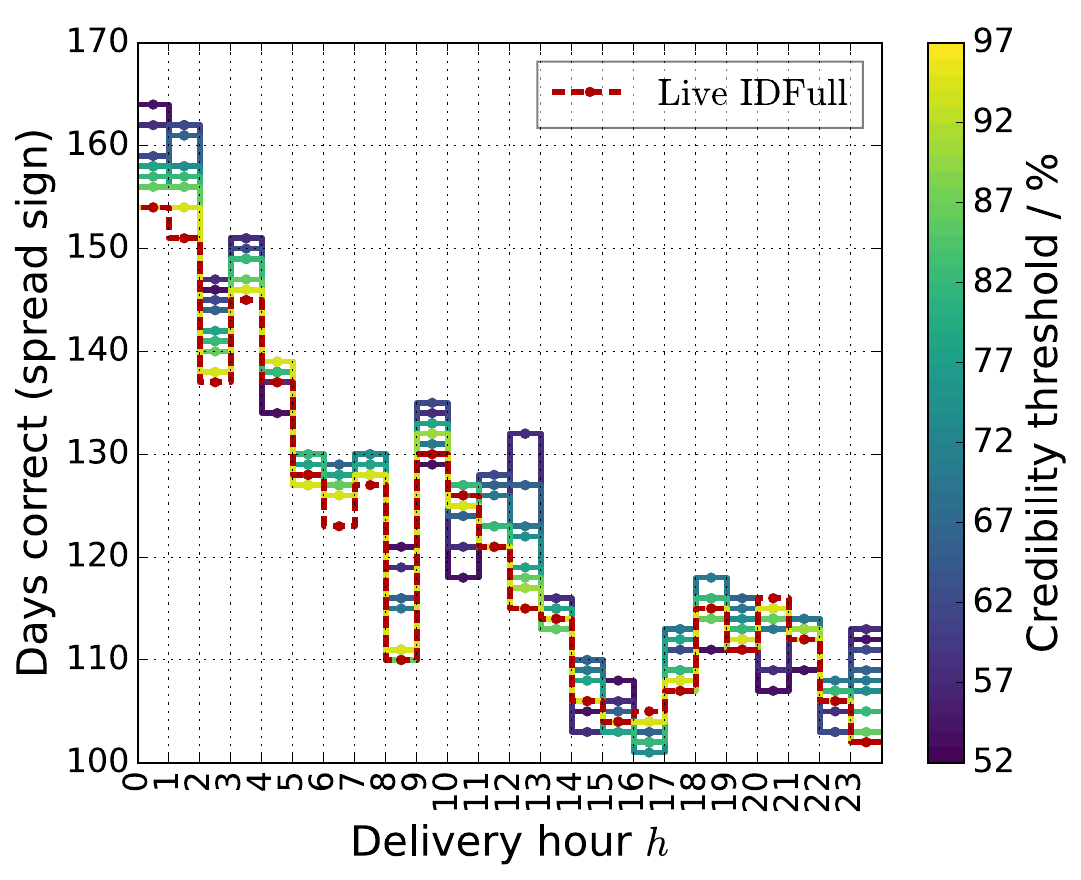} \hfil \includegraphics[width=0.45\textwidth]{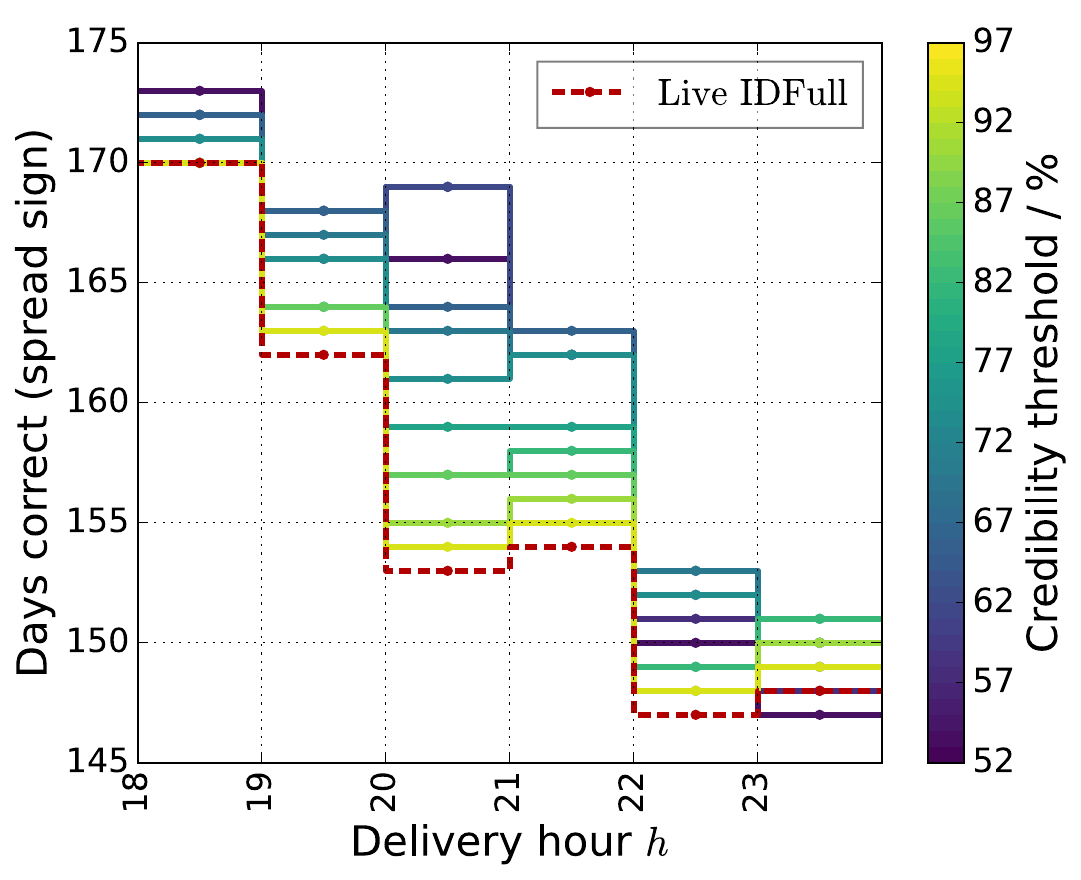}
	\end{center}
	\phantom{X}{Scenario (e), $\hlag=1$} \hfil \phantom{X}{Scenario (e), $\hlag=5$} \hfil\\[-7mm]
	\begin{center}
	\includegraphics[width=0.45\textwidth]{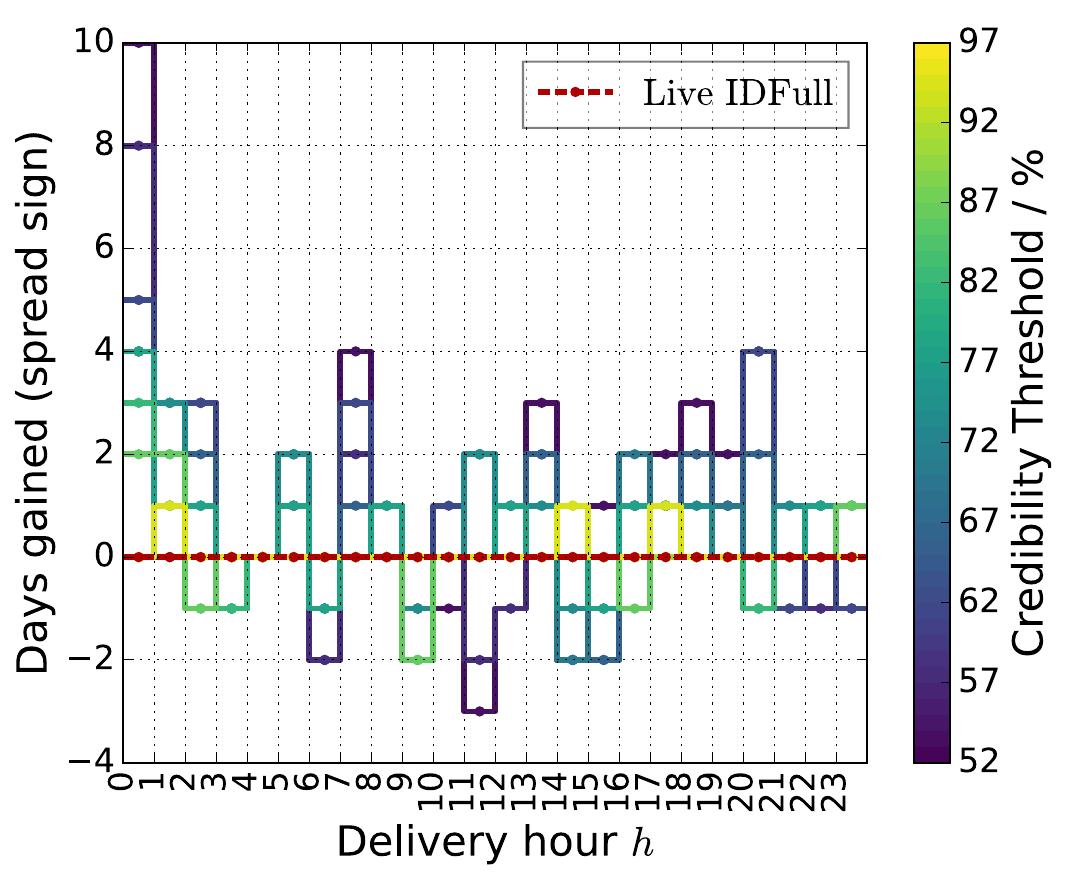} \hfil \includegraphics[width=0.45\textwidth]{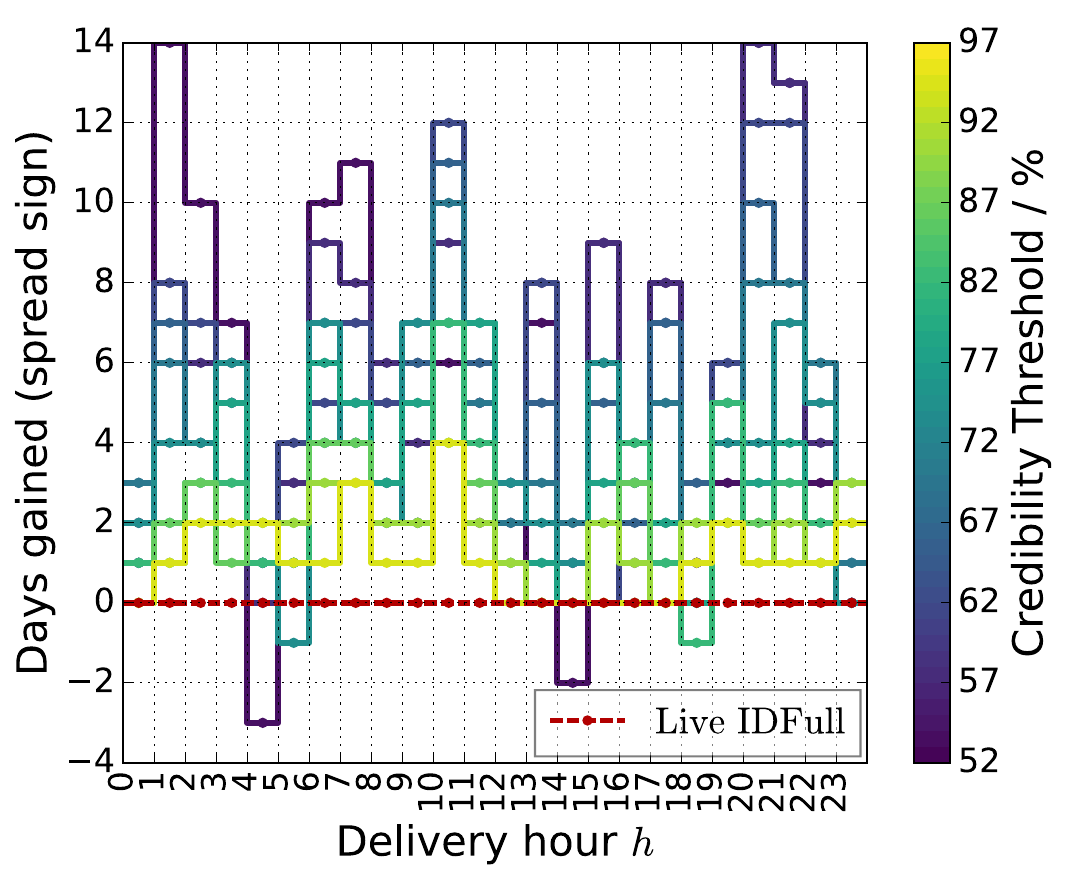}
	\end{center}
	\caption{Exemplary forecast results of the sign spread using the estimator in \eqref{eq:sgn_spread}. The top panel shows the number of days with correct forecasts for the two scenarios (a) and (d) with fixed forecast creation time $\tau$, where the dashed red line shows the reference result using $\mr{sgn}(P_\mr{idfull})$. The bottom panel shows two examples of scenario (e) with fixed lag $\hlag=1$ and $\hlag=5$. Here, the IDFull reference has been subtracted from the forecast results, such that the number of correctly forecasts days gained using \eqref{eq:sgn_spread} are obtained. The colour code represents the choice of the credibility threshold $p_0$. The total number of days considered is 183.}
	\label{fig:days}
\end{figure*}

In Figure \ref{fig:days} we show examples of counts of days with correct spread sign forecasts $\hat s(\mr{spread})$ for different values of the credibility threshold $p_0$. As expected, when $p_0$ approaches $1$, the estimator $\hat s(\mr{spread})$ reduces to $\mr{sgn}(P_\mr{idfull})$. For values of $p_0$ close to 0.5, the number of days with correctly forecasted spread sign tend to be maximal, but also the risk of a forecast inferior to the live IDFull reference increases. A sweet spot between $p=0.5$ and $p=1$ may exist, but is not apparent in this study and will be left for future work.

\begin{figure*}[t!]
	\begin{center}
	\includegraphics[width=0.45\textwidth]{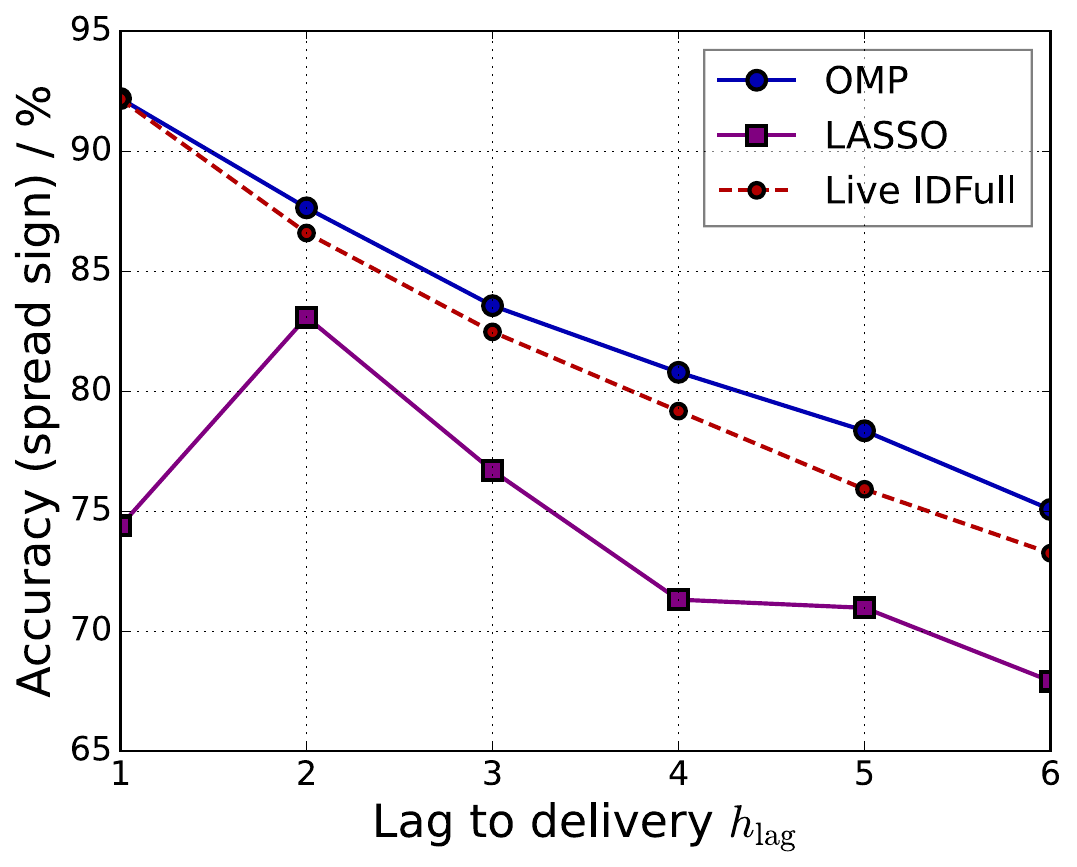} \hfil \includegraphics[width=0.45\textwidth]{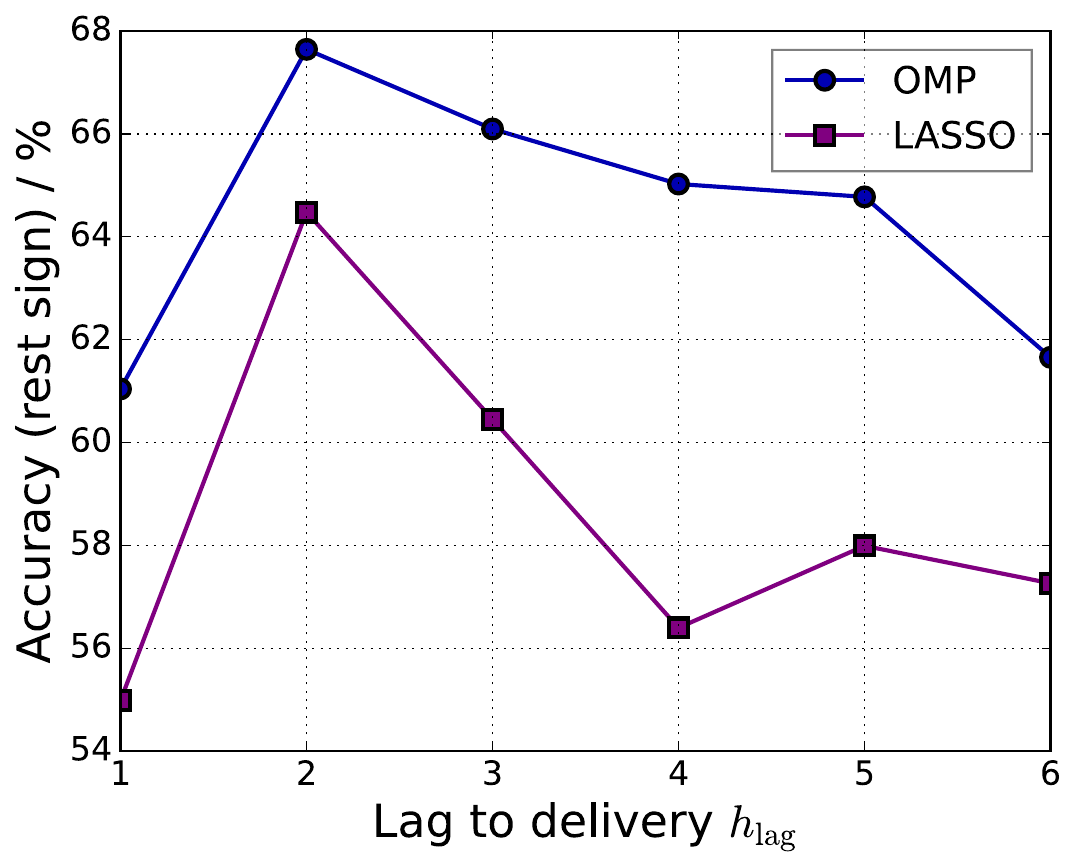}
	\end{center}
	\caption{The accuracy of the estimators \eqref{eq:sgn_spread} and \eqref{eq:sgn_rest} for the sign of the \textit{spread} and the \textit{rest} are shown, together with the IDFull reference for the spread sign, cf. Figure \ref{fig:days}. Here, the scenarios (e) and (f) investigate the difference in performance using OMP and LASSO for feature selection. OMP demonstrates an average increase of $1.7\,\%$ in spread sign forecast accuracy compared to the live benchmark, and increases of $12.1\,\%$ and $10.0\,\%$ for spread and rest sign forecast accuracy compared to LASSO. LASSO leads to a decrease in accuracy by $-9.0\,\%$ compared to the live benchmark. The accuracy is determined as number of correctly forecasted days across all hours over the total number of forecasts.}
	\label{fig:days_overall}
\end{figure*}

For an overall comparison of the sign forecast accuracy for both the spread and the rest, we investigate the choice $p_0=0.5$ for the fixed lag scenarios (e) and (f) in Figure \ref{fig:days_overall}, where the accuracy as the ratio of correct forecasts over total number of forecasts is shown. The forecasts using OMP feature selection results in a considerably higher accuracy than the forecasts using LASSO feature selection, and also turns out to be better than the live benchmark $\mr{sgn}(P_\mr{idfull})$. Later we show that these results are statistically significant.

\subsection{Probabilistic forecast scores}
In evaluating probabilistic forecasting, the main difference to point estimates is that a true predictive distribution to compare against is not available, instead, the estimated predictive distribution $\pde(\y)$ can only be compared against the true price $\ypr=\mr{eod}_d(P_\mr{idfull})$. 

The review \citet{Maciejowska_forecasting_2023} discusses the evaluation of probabilistic forecasting in the context of EPF. Two main aspects are relevant here, which are reliability and sharpness. Reliability describes how well forecast uncertainty is captured by the probabilistic nature of the forecast. For instance, if the PIs that carry 95\% of the probability mass cover the true value in 95\% of all forecasts produced by a method, then this method would be perfectly reliable for a credibility level of $\alpha=0.95$. Sharpness refers to the accuracy of the forecast, in the sense that PIs with small width but high credibility provide more localised estimates.

To assess these two forecast qualities, quantitative scoring rules are gaining popularity in the probabilistic EPF forecasting literature \citep{Maciejowska_forecasting_2023}. These scores are based on PIs extracted from predictive distributions, as well as on the predictive distributions itself.

\begin{figure*}[t!]
	\begin{center}
	\includegraphics[width=0.45\textwidth]{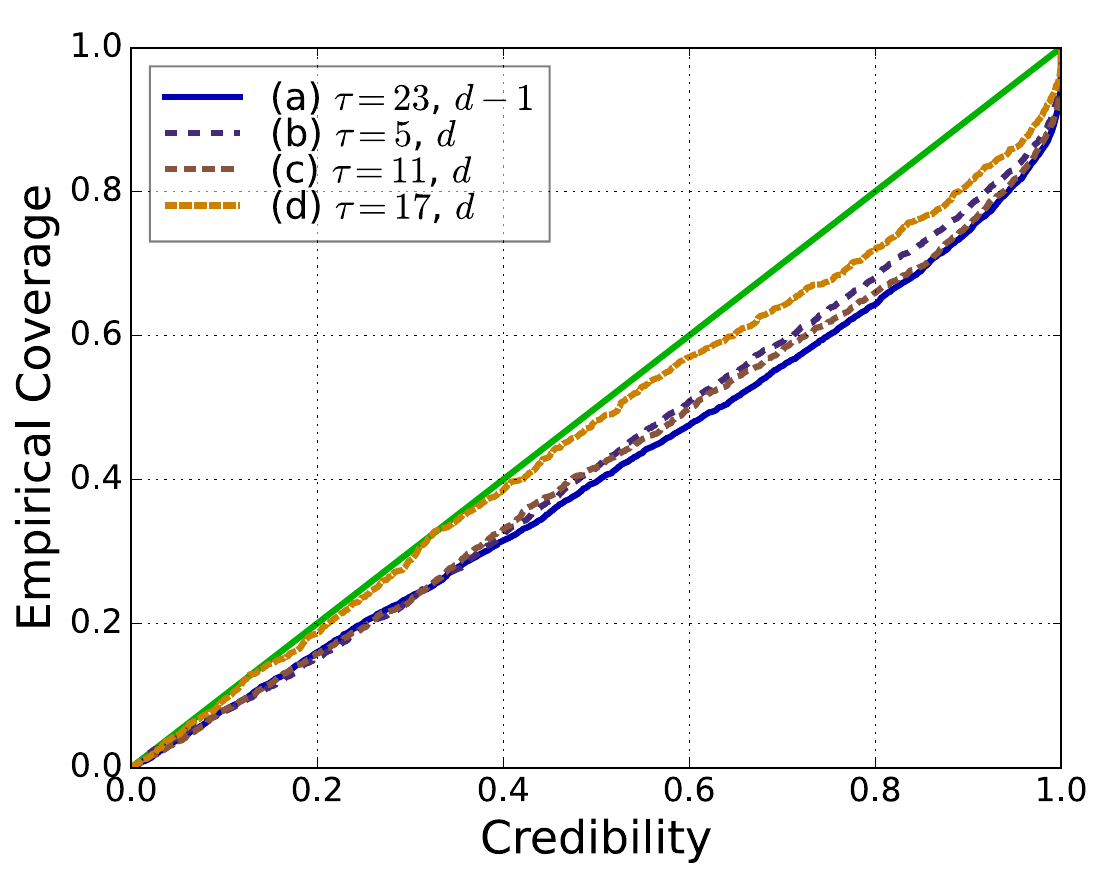} \hfil \includegraphics[width=0.45\textwidth]{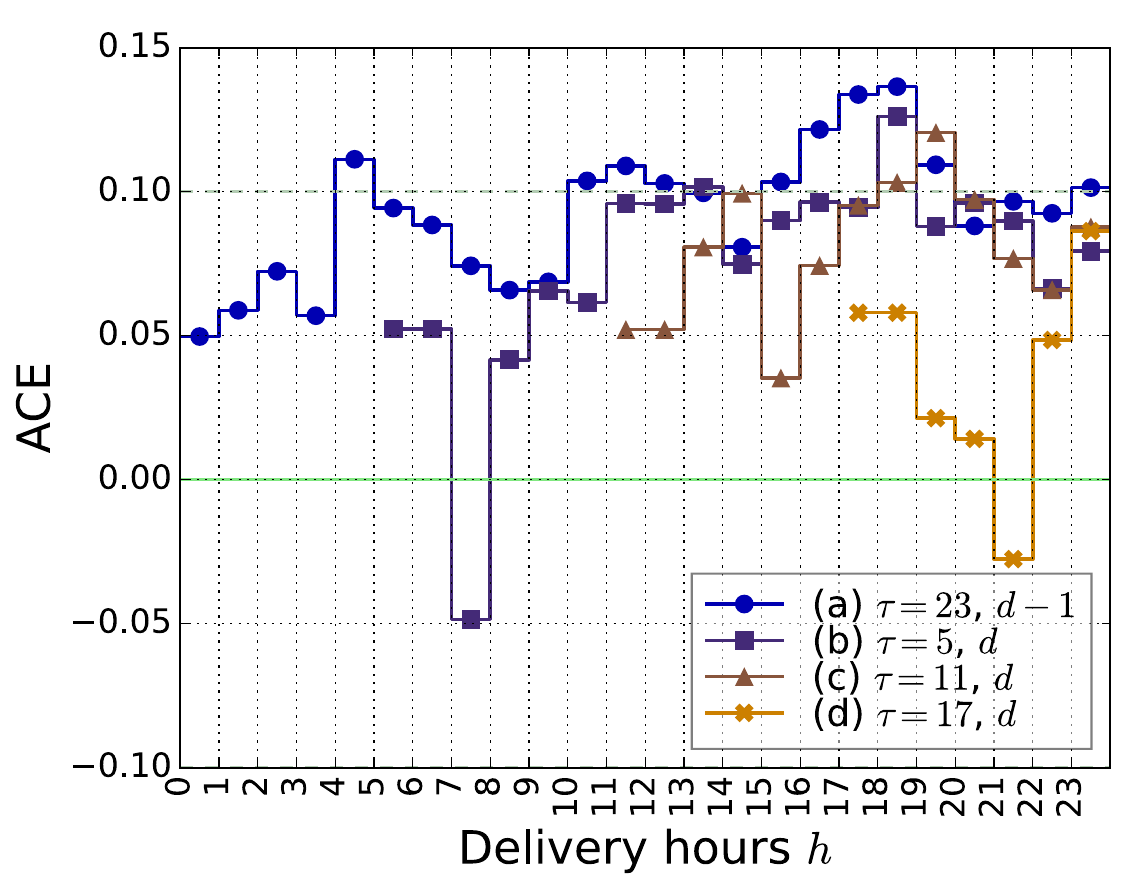}
	\end{center}
	\caption{The Empirical Coverage (left) and the resulting Average Coverage Error (ACE) (right) are shown for the fixed creation times $\tau$ of scenarios (a) - (d). The solid green lines represent the theoretical expectation in case of perfect coverage.}
	\label{fig:empicov_ace_fixcreat}
\end{figure*}	

\begin{figure*}[t!]
	\phantom{X}{Scenario (e) \;(OMP feature selection)} \hfil \phantom{X}{Scenario (f) \;(LASSO feature selection)} \hfil\\[-7mm]
	\begin{center}
	\includegraphics[width=0.45\textwidth]{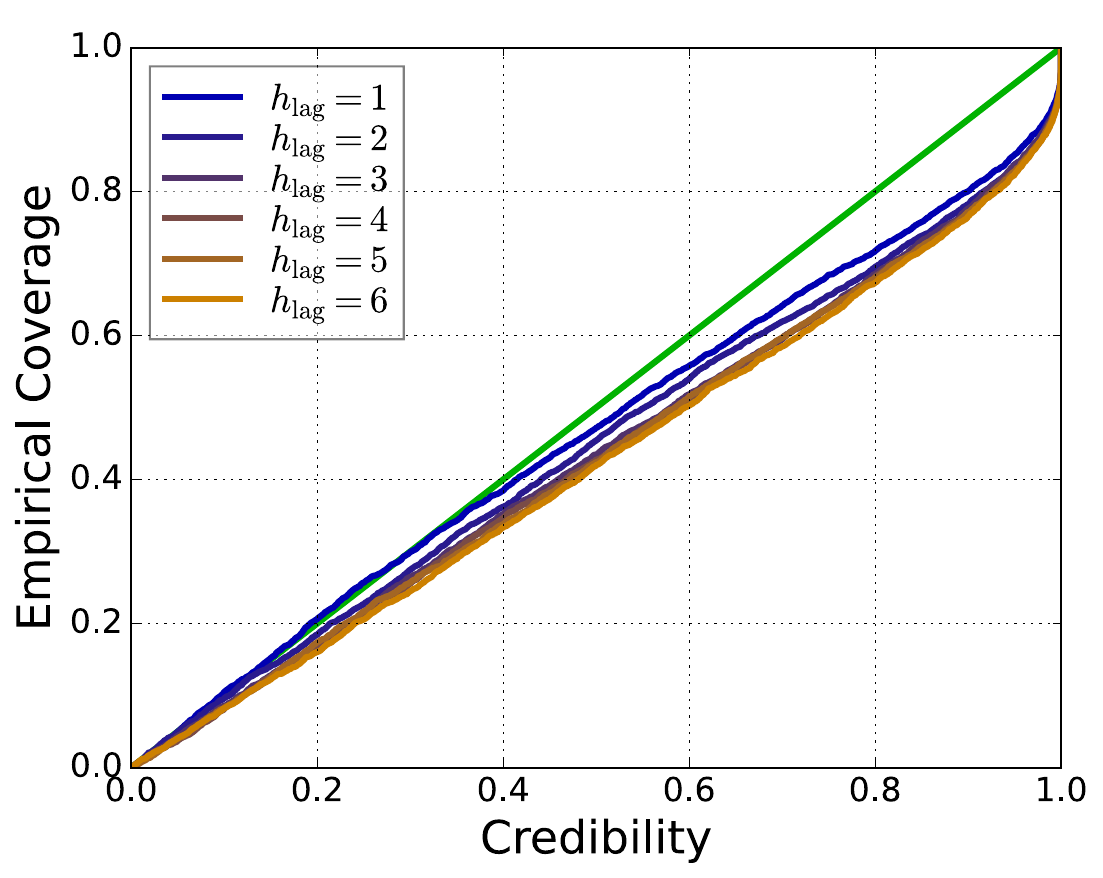} \hfil \includegraphics[width=0.45\textwidth]{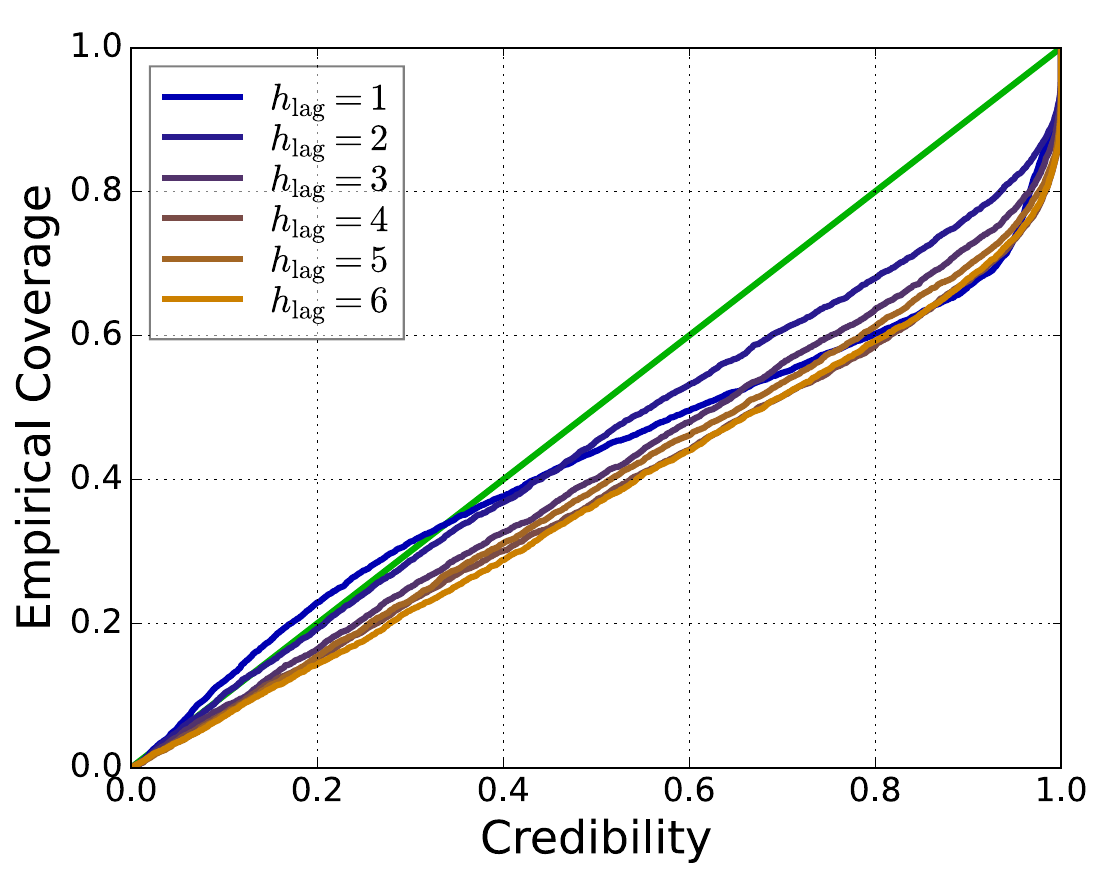}
	\end{center}
	\phantom{X}{Scenario (e) \;(OMP feature selection)} \hfil \phantom{X}{Scenario (f) \;(LASSO feature selection)} \hfil\\[-7mm]
	\begin{center}
	\includegraphics[width=0.45\textwidth]{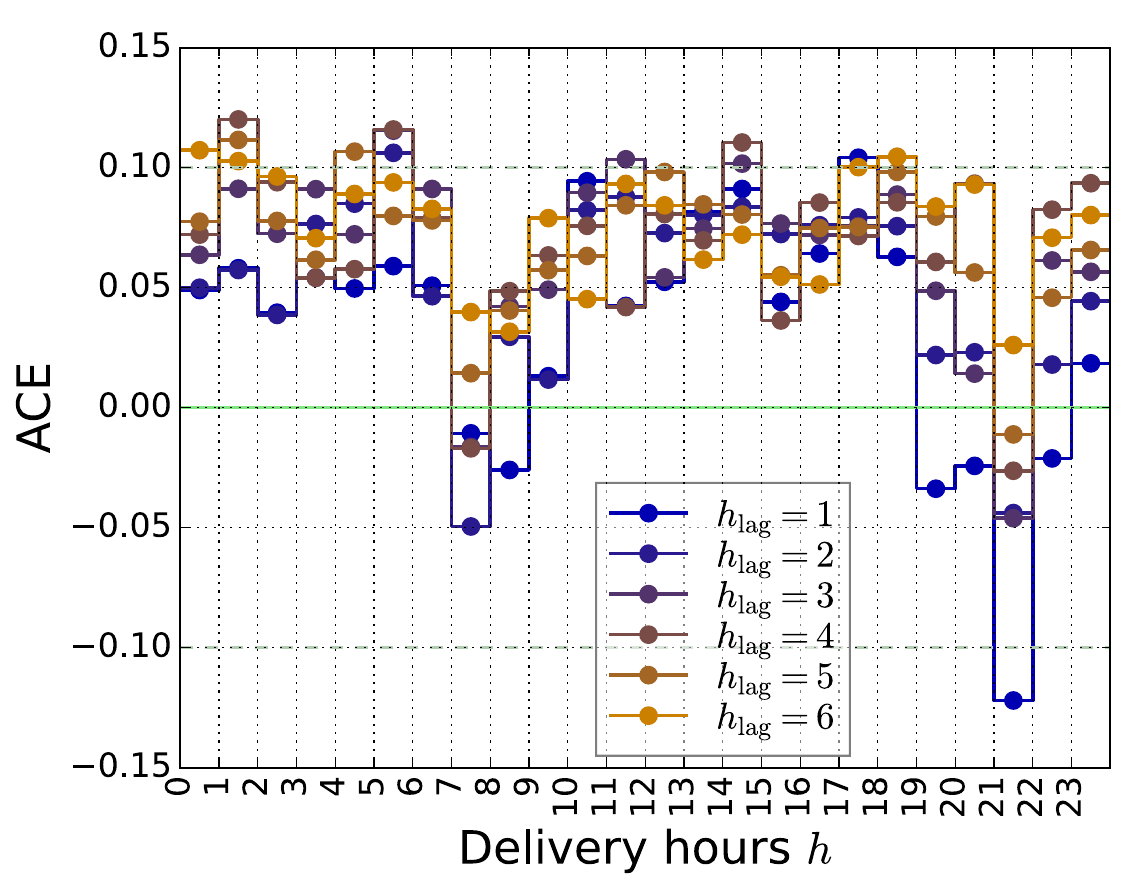} \hfil \includegraphics[width=0.45\textwidth]{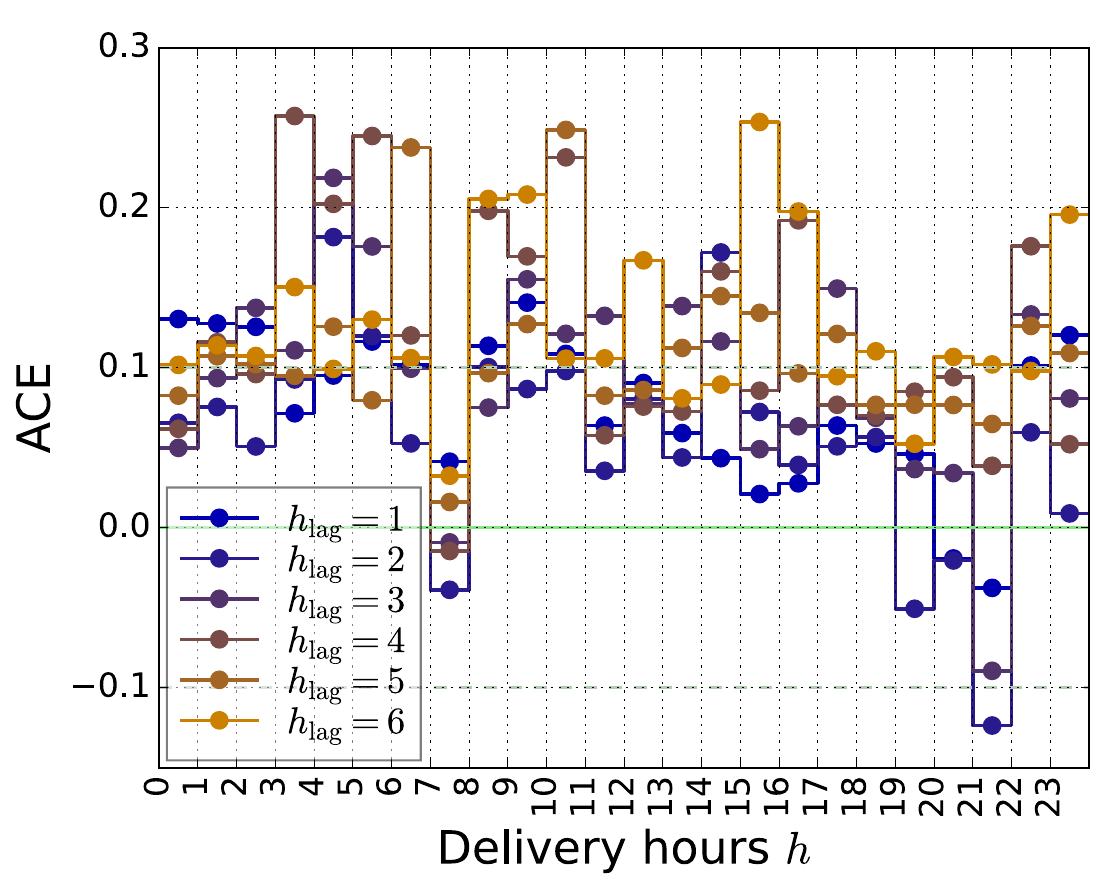}
	\end{center}
	\caption{The Empirical Coverage (top panel) and the resulting Average Coverage Error (ACE) (bottom panel) are shown for scenarios (e) and (f) with fixed lag between forecast creation and delivery begin. The left panel uses OMP feature selection, the right panel LASSO feature selection. The solid green lines represent the theoretical expectation in case of perfect coverage.}
	\label{fig:empicov_ace_fixlag}
\end{figure*}

\begin{figure*}[t!]
	\begin{center}
		\includegraphics[width=0.45\textwidth]{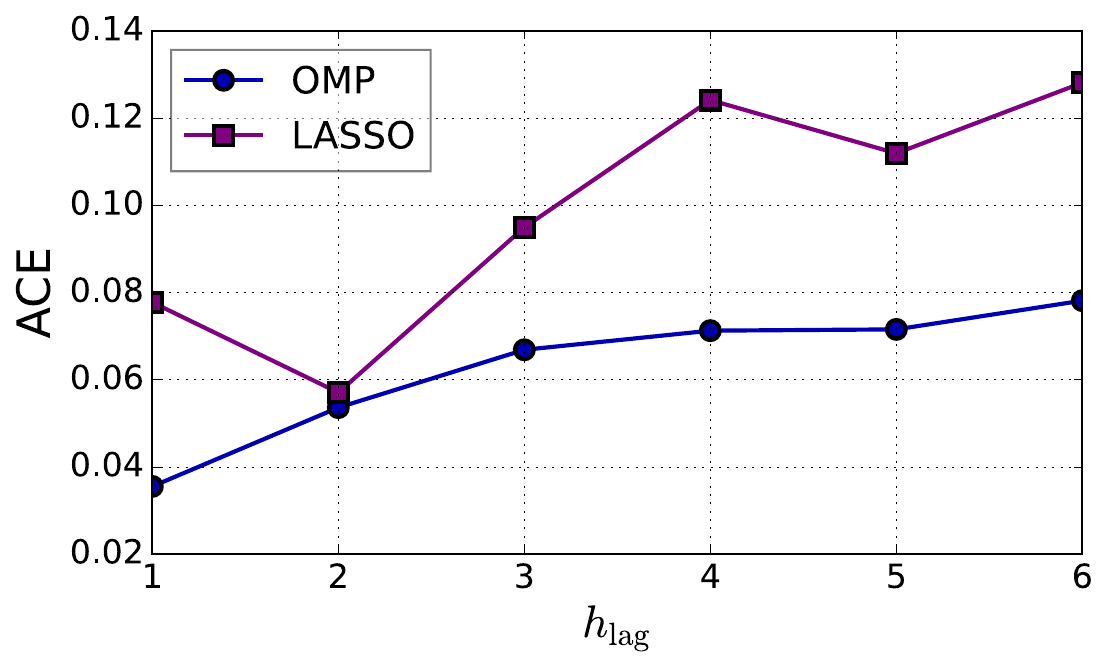}\hfil \includegraphics[width=0.45\textwidth]{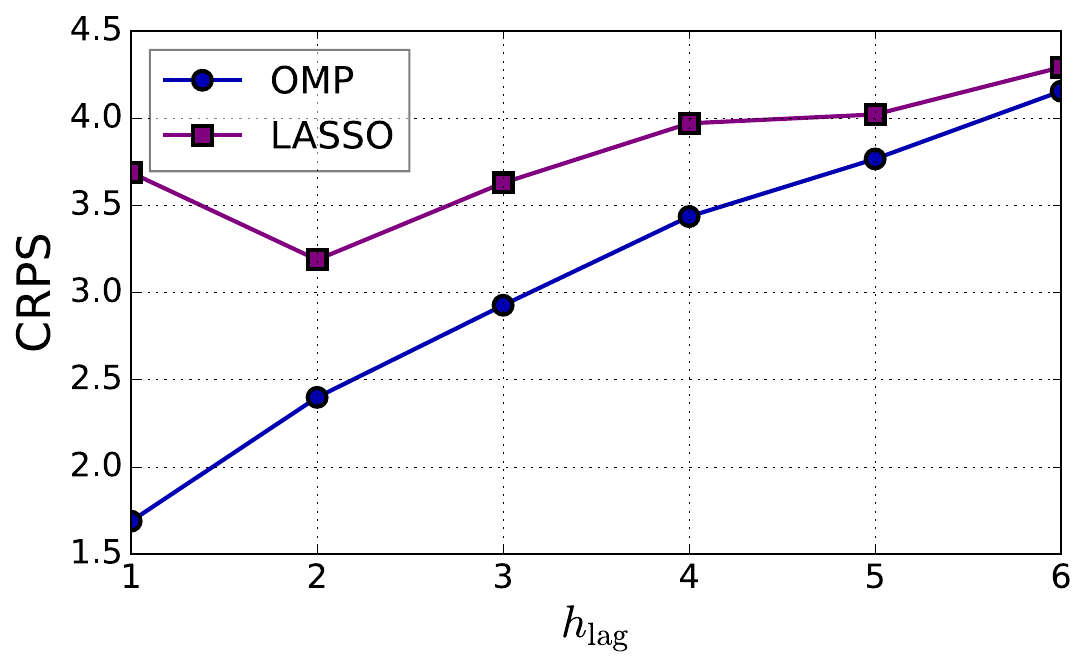}\\
	\end{center}
	\caption{Comparison between the use of OMP and LASSO for feature selection for scenarios (e) and (f) averaged across all days and hours of the test period. On the left, we show the Average Coverage Error (ACE), with OMP showing an average $34.59\,\%$ decrease compared to LASSO. On the right, we show the Continuous Ranked Probability Score (CRPS) defined in  \eqref{eq:crps}, with OMP showing an average $20.21\,\%$ decrease compared to LASSO.}
	\label{fig:ace_cprs}
\end{figure*}

To directly target reliability, we test the empirical coverage by counting how often the true $\mr{eod}_d(P_\mr{idfull})$ was contained in a PI at given credibility $\alpha$. Normalizing and plotting against $\alpha$, we can visually inspect how closely the empirical coverage follows the theoretical expectation given by the unit diagonal. In Figures \ref{fig:empicov_ace_fixcreat} and \ref{fig:empicov_ace_fixlag} we depict the empirical coverage in the left panels for all scenarios. Averaging the deviation from the theoretical diagonal, we obtain the Average Coverage Error (ACE), which is depicted on the right panels of the same Figures \ref{fig:empicov_ace_fixcreat} and \ref{fig:empicov_ace_fixlag}. It can be seen that all forecasts tend to be a little overconfident towards higher credibilities, a signature of the highly volatile prices on the German CID market. Using OMP instead of LASSO reduces the ACE by a factor of about 2. An overall comparison between OMP and LASSO in terms of ACE is shown on the left in Figure \ref{fig:ace_cprs}.

To test both reliability and sharpness, the Continuous Ranked Probability Score (CRPS) can be used \citep{gneiting_strictly_2007}, 
\begin{align}
\crps(\pde,\ypr) = \Ex[|\Y-\ypr|] - \frac{1}{2}\Ex[|\Y_1-\Y_2|] \,. \label{eq:crps}
\end{align}
Here, it is assumed that $\Y,\,\Y_1,\,\Y_2\overset{iid}{\sim} \pde$, and the expectations are taken with respect to $\pde$. With the full posterior predictive distribution available as an interpolation object using \texttt{scipy.interpolate} in \textsc{python} \citep{SciPy-NMeth}, we compute the expectation values to practically arbitrary precision directly from their definitions in term of integrals over $\pde$,
\begin{align}
\Ex[|\Y-\ypr|] &= \int_{-\infty}^{\infty}  |\y'-\ypr| \, \pde(\y') \,\d \y' 
\end{align}
and
\begin{align}
\Ex[|\Y_1-\Y_2|] &= \int_{-\infty}^{\infty} |z|\, \rho_Z(z) \,\d z \,,
\end{align}
with the density $\rho_Z(z)$ for $Z=\Y_1-\Y_2$ given by
\begin{align}
\rho_Z(z) = \int_{-\infty}^{\infty} \pd(\y) \, \pd(\y-z) \,\d \y \,.
\end{align}

We compute the CRPS for each forecast of scenarios (e) and (f) and show the result on the right in Figure \ref{fig:ace_cprs}. Again, OMP turns out to deliver better results than LASSO, which is statistically significant as will be shown in the next section.

\subsection{Statistical significance of results}
In all evaluations, the forecasts making use of OMP for feature selection perform better than using LASSO for feature selection, and also better than the live IDFull benchmark. To investigate the statistical significance of this overall result, we employ the Diebold-Mariano (DM) test \citep{diebold_comparing_1994}. For the scenarios (e) and (f) with fixed lag $\hlag$, the DM test is basically a $z$-test for the difference of mean between scores of two forecast series'. The test is agnostic with respect to the choice of score, as long as the score can be computed for individual forecasts. Here, the MAE, a Boolean variable reflecting correct sign forecasts and the CRPS can serve as scores for the DM test. We employ the \textsc{python} package \texttt{dieboldmariano} which uses the Harvey correction generalizing the test statistics to the standard $t$-distribution to account for smaller sample sizes \citep{harvey_testing_1997}. 

\begin{table*}[h!]
	\begin{center}
		\includegraphics[width=0.75\textwidth]{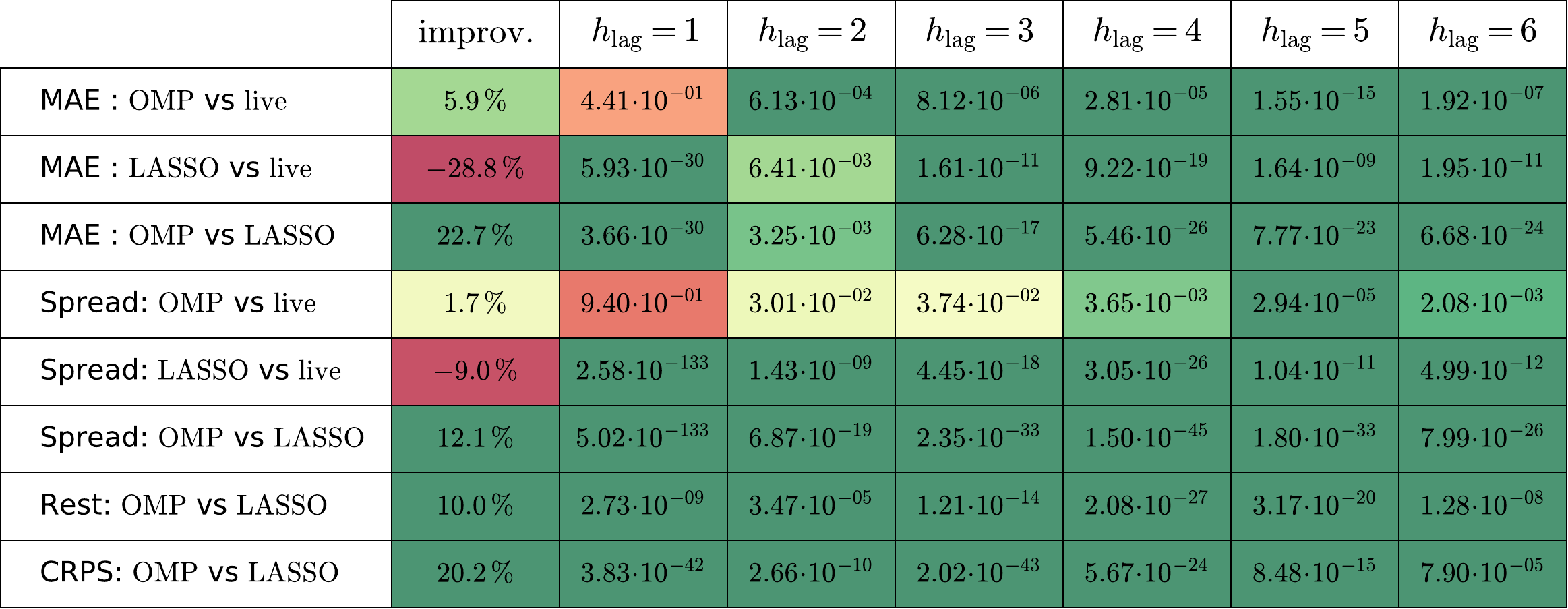} 
	\end{center}
	\caption{The average improvements and corresponding $p$-values obtained by applying the one-sided Diebold-Mariano (DM) test are listed for various scores and forecast series'. As scores we consider the MAEs shown in the top right chart of Figure \ref{fig:MAE_abcdef}, a Boolean variable reflecting true sign forecasts of the \textit{spread} and the \textit{rest} as used for Figure \ref{fig:days_overall}, and the CRPS shown in Figure \ref{fig:ace_cprs} on the right. The forecast series' tested are those of scenarios (e) and (f), that is using OMP and LASSO for feature selection, as well as the live IDFull benchmark. Strong statistical significance and improved forecast performance are indicated by green colours, no statistical significance and declined forecast performance by red colours.}
	\label{tab:signi_fixlag}
\end{table*}

We test the one-sided null hypothesis that all forecast series of scenarios (e) and (f) are statistically identical in terms of the MAE, the spread and rest sign, and the CRPS. The resulting $p$-values are shown in Table \ref{tab:signi_fixlag}. Apart from very few examples that coincide with cases where the scores appear almost indistinguishable, we find vanishing $p$-values and thus confirm the statistical significance of the observations made in this work.

\section{Conclusions}\label{sec:conclusion}
We presented a forecast study of the IDFull electricity price index of the German continuous intraday market for the years 2021 and 2022. These years are characterised by unprecedented volatility and have hardly been studied before. We provided details about reproducing price indices and statistics published by EPEX Spot using transaction data of the continuous intraday market provided by \citet{EPEXdata}. 

Due to the strong volatility of intraday prices in 2021 and 2022, it is mandatory to employ probabilistic forecasting. We demonstrated how Bayesian models can successfully be deployed to obtain posterior predictive distributions fully incorporating parameter uncertainty. We further presented how point estimates, predictive intervals and forecast probabilities can be extracted from the predictive distributions.

A currently debated topic in the literature is the supposed weak-form efficiency of the continuous intraday market. According to this hypothesis, all information available is already contained in last prices due to informed traders, making it impossible to significantly beat last price information by means of forecasting models. In our study, we partly confirm the hypothesis in that our model clearly identifies current price information as the dominating regressor and closely follows its trend. But on the other hand, we find that the live IDFull built from current prices can still be improved in a statistically significant way. It should, however, be taken into consideration that the definition of a last-price benchmark is not unique, and other last price information, or combinations therefore, might still deliver the best forecast possible. It should also be noted that the continuous intraday market is developing rapidly, making it difficult for traders to adjust to changes, which in turn opens up the potential of forecast models to beat last-price benchmarks. 

Our conclusion on this debate from this study therefore is that the weak-form efficiency can tentatively be confirmed in the sense of a solid characterisation of market properties and possible future developments, but comprehensive forecast models may still be able to beat last-price benchmarks. Aside from the question of weak-form efficiency of markets, last-price benchmarks will still be point estimates, and even a probabilistic forecast that does not improve a last-price benchmark but yields reliable uncertainty distributions around the benchmark is a fruitful directions of research and would be of great value for traders.

Another aspect discussed in the literature that we address in our study is feature selection. Comprehensive forecast models that potentially outperform last-price benchmarks often draw from a large pool of regressors. These variables typically exhibit strong collinearities, which impede robust feature selection. Previous works have found that LASSO is effective for feature selection and is thus considered the gold standard. However, LASSO can still exhibit instabilities under strong collinearities. A more robust alternative is Orthogonal Matching Pursuit (OMP), which we suggest as an alternative to LASSO. In our study, we find a clear improvement using OMP instead of LASSO, with strong statistical significance.

In summary, the innovative contribution of this work to electricity price forecasting primarily lies in the Bayesian processing of full parameter uncertainty information rather than reducing it to point estimates, along with the probabilistic analysis of the resulting posterior predictive distributions. Additional contributions include the reproduction and live values of all intraday price indices and statistics, and handling the large set of strongly correlated features with OMP and a regularising prior. Lastly, we advocate for probabilistic modelling using \textsc{TensorFlow Probability}.

Our model can be developed further in various ways offering plenty possibilities for future research, such as systematically exploring different error distributions with long tails and skewness, adding non-linearities to the basic regression approach, and employing ensemble or mixture models. 

Overall, with our work, we hope to strengthen the field of probabilistic electricity price forecasting, promoting the use of Bayesian models that can fully incorporate parameter uncertainty.

\section{Acknowledgements}
We gratefully acknowledge fruitful discussions with Sebastian Uhl. We thank Sovann Khou from EPEX SPOT for providing additional information on the calculation of price indices and statistics. 




\bibliographystyle{elsarticle-harv} 
\bibliography{recent_electricty_papers.bib,code.bib,statistical_learning.bib}




\end{document}